# Buildup and dephasing of Floquet-Bloch bands on subcycle time scales

S. Ito[1,†], M. Schüler[2,†], M. Meierhofer[3], S. Schlauderer[3], J. Freudenstein[3], J. Reimann[1], D. Afanasiev[3],

K. A. Kokh[4], O. E. Tereshchenko[4], J. Güdde[1], M. A. Sentef[5,*], U. Höfer[1,3,*], R. Huber[3,*]

[1]Department of Physics, Philipps-University of Marburg, 35032 Marburg, Germany

[2]Condensed Matter Theory Group, Paul Scherrer Institute, CH-5232 Villigen PSI, Switzerland

[3]Department of Physics, University of Regensburg, 93040 Regensburg, Germany

[4]A.V. Rzhanov Institute of Semiconductor Physics and V.S. Sobolev Institute of Geology and Mineralogy

SB RAS, 630090, Novosibirsk, Russian Federation

[5]Max Planck Institute for the Structure and Dynamics of Matter, 22761 Hamburg, Germany

†These authors contributed equally, *e-mail: michael.sentef@mpsd.mpg.de; hoefer@physik.uni-

marburg.de; rupert.huber@physik.uni-regensburg.de

**Strong light fields have created spectacular opportunities to tailor novel functionalities of solids[1-5]. Floquet-Bloch states can form under periodic driving of electrons and enable exotic quantum phases[6-15]. On subcycle time scales, lightwaves can simultaneously drive intraband currents[16-29] and interband transitions[18,19,30,31], which enable high-harmonic generation[16,18,19,21,22,25,28-30] (HHG) and pave the way towards ultrafast electronics. Yet, the interplay of intra- and interband excitations as well as their relation with Floquet physics have been key open questions as dynamical aspects of Floquet states have remained elusive. Here we provide this pivotal link by pioneering the ultrafast buildup of Floquet-Bloch bands with time- and angle-resolved photo-emission spectroscopy. We drive surface states on a topological insulator[32,33] with mid-infrared fields – strong enough for HHG – and directly monitor the transient band structure with subcycle time resolution. Starting with strong intraband currents, we observe how Floquet sidebands emerge within a single optical cycle; intraband acceleration simultaneously proceeds in multiple sidebands until high-energy electrons scatter into bulk states and dissipation destroys the Floquet bands. Quantum nonequilibrium calculations explain the simultaneous occurrence of Floquet states with intra- and interband dynamics. Our joint experiment-theory study opens up a direct time-domain view of Floquet physics and explores the fundamental frontiers of ultrafast band-structure engineering.**

In a crystalline solid, the discrete translational symmetry of the ionic potential implies that the band structure of Bloch electrons is periodic in momentum space. Analogously, a time-periodic potential of light with frequency $\Omega$ acting on a Bloch electron replicates its band structure as 'Floquet' copies[1,4,5] shifted in energy by multiples of $\hbar\Omega$. When the lightwave potential approaches the strength of the static crystalline one, Floquet-Bloch sidebands hybridize among each other[1,6-10] and with ground-state bands[11], shaping a quasi-stationary band manifold under periodic driving (Fig. 1a). Floquet engineering has raised high hopes to create novel quantum effects, such as dynamical localization[5], photon-dressed topological states[6-10], and time crystals[12,13]. Besides proof-of-concept experiments in atomic systems[12,13], quasi-stationary Floquet-Bloch states in topological insulators have been studied[7,8], and emergent quantum phenomena such as a light-induced anomalous Hall effect in graphene[14] and magnonic space-time crystals have been observed[15]. Yet, the dynamics of Floquet states, such as their transient emergence and the transfer of electrons between sidebands[34,35], has not been experimentally accessible.

Meanwhile, the interaction of atomically strong light fields with Bloch electrons has been explored on time scales shorter than a single cycle of light[18-24,26-31]. The carrier wave has been shown to ballistically accelerate electrons through individual bands[16-29] and to simultaneously induce non-perturbative multi-photon interband transitions[18,19,30] (Fig. 1a). The resulting dynamics have been utilized for HHG[16,18,19,21,22,25,28-30], electron-hole recollisions[17,20,27], lightwave valleytronics[26] or carrier-envelope-phase-controlled photocurrents[23]. While intraband currents driven by relatively weak terahertz fields have been visualized in angle-resolved photoemission spectroscopy (ARPES)[24], the description of strong-field interband excitations and their interplay with intraband dynamics has relied on theoretical models. More generally, the connection between subcycle dynamics and Floquet states associated with multicycle periodic driving has remained largely elusive despite its crucial role in strong-field light-matter interaction and its potential for shaping dynamic phases of matter by light.

Here we directly visualize the ultrafast buildup and dephasing of a Floquet-Bloch band structure with subcycle ARPES. The first half-cycle of an intense mid-infrared (MIR) lightwave accelerates electrons within the Bloch band structure before Floquet sidebands emerge within a single optical cycle. Simultaneously, intraband acceleration proceeds within multiple sidebands, until high-energy electrons scatter into bulk states and dissipation destroys the Floquet bands. Our lightwave ARPES observations are in line with quantum-nonequilibrium calculations and establish the ultrafast limit of optical band structure design.

**Strong-field band-structure videography**

The three-dimensional topological insulator $Bi_2Te_3$ forms an ideal platform for our study. This material features a finite band gap in the bulk and a Dirac-like topological surface state (TSS) (Fig. 1b). Spin-momentum locking[32,33] protects the TSS from scattering[24], which warrants sufficient coherence times for Floquet-Bloch states to emerge under periodic driving[7,8,36]. We excite the system with intense, phase-locked, s-polarized MIR pulses with a centre frequency of $\Omega_{MIR}/2\pi = 25$ THz (photon energy, $\hbar\Omega_{MIR} = 0.1$ eV) (Fig. 1b). Simultaneously, the ultrafast evolution of the band structure and its occupation are mapped out by lightwave ARPES: an ultraviolet probe pulse arriving at a variable delay time, $t$, photoemits electrons by a two-photon process; the energy and momentum of the photoelectrons are detected with a hemispherical electron analyser. For a time resolution better than half an oscillation period of the MIR driving field, we compress the photoemission probe to 17 fs (Extended Data Fig. 1), which inevitably leads to spectral broadening of the ARPES data. The visibility of fine band structure details on the energy scale of $\hbar\Omega_{MIR}$ can be augmented by second-derivative image processing, which is well established in high-resolution ARPES[37] to precisely reconstruct the original peak positions with the desired sensitivity (Extended Data Fig. 2).

The intense MIR field not only drives dynamics inside the sample but also interacts with emitted photoelectrons *in vacuo*[24]. P-polarized MIR light induces well-defined displacements of the whole projected band map along the energy axis in a process called energy streaking[24], whereas s-polarized light shakes the momentum coordinates. Such momentum streaking allows for a quantitative *in-situ* reconstruction of the electric-field waveform directly on the sample surface, which is difficult to access otherwise[24]. The procedure works well also for curvature-filtered ARPES spectra (Extended Data Fig. 3). The valence band streaking trace (Fig. 2a) is readily converted into an s-polarized electric-field waveform (Fig. 2b), evidencing the subcycle resolution of our setup at MIR frequencies. The retrieved peak field of 0.8 MV cm$^{-1}$ corresponds to an incident amplitude of $\sim 7$ MV cm$^{-1}$ (see Methods) and is strong enough for HHG in the TSS[28].

**Buildup of Floquet-Bloch sidebands**

Streaking compensated ARPES maps[24] reveal the material-inherent dynamics of the band structure (Fig. 2c). Up to 100 fs before the maximum field crest, the effect of the MIR lightwave on the TSS is negligible. The ARPES image shows a partially filled Dirac cone, in agreement with our DFT calculations for $Bi_2Te_3$. Forty fs later ($t$ = -60 fs), the electric field has induced a strong asymmetry in the electronic distribution along the momentum axis – the hallmark of lightwave-driven currents. Most remarkably, a distinct splitting of electronic populations into multiple branches becomes noticeable only one quarter of an oscillation cycle later, at $t$ = -50 fs. Simultaneously, the electron population oscillates back and forth along the momentum axis while the driving field evolves (Fig. 2c, $t$ = -50 fs and -40 fs). Figure 2b (red circles) displays the temporal evolution of the spectrally integrated current $j_y$ retrieved by curvature filtering with less peak sensitivity (Extended Data Fig. 3). Even under the extreme conditions of our strong-field MIR lightwaves, the total subcycle current is well described by the semiclassical Boltzmann equation (Fig. 2b, red curve), treating electrons like Dirac particles. Yet, the emergent band splitting indicates quantum coherences beyond any semiclassical model.

The split bands notably follow the ground-state band structure shifted by integer multiples of $\hbar\Omega_{MIR}$ = 0.1 eV (Fig. 2c, thin grey curves). The visibility of the spectral modulations is less pronounced at lower fields with all other features remaining intact (Extended Data Fig. 4). As the field increases, the band splitting becomes more prominent and the probability to find the electrons shifts entirely to the upper branch near the field crest (Fig. 2c, $t$ = -18 fs), indicating the signatures of Floquet-Bloch states. This observation is surprising because the conventional Floquet picture assumes multi-cycle lightwaves to form sidebands by coherent superposition of electronic states[7,8], whereas our sub-cycle experiments identify Floquet-Bloch states already within the first two cycles of a MIR pulse. We can exclude alternative effects to cause these features: potential artefacts of the curvature analysis can be ruled out because the band splitting is well resolved in the raw ARPES spectra (Extended Data Figs. 5 and 6). So-called laser-assisted photoemission (LAPE)[38] or Volkov states[8,36] created by the interaction of the emitted photoelectrons with the MIR lightwave have been shown to be strongly suppressed when the driving field is polarized in the surface plane[8], which is our configuration (Fig. 1b). Indeed, lightwave

ARPES maps recorded for a non-zero MIR electric field component perpendicular to the surface differ significantly from Fig. 2c (Extended Data Fig. 7). We also verified that the two-photon probing is performed under non-resonant conditions and thus cannot introduce sideband artefacts (Extended Data Fig. 8). Note that hybridization between Floquet-Bloch sidebands is known to modify their dispersion and open small gaps[7,8]. The relevant dispersion direction, however, is perpendicular to the polarization vector of the driving electric field and not parallel as in our experimental setup.

To understand how MIR-driven intraband acceleration leads to the emergence of Floquet sidebands already in the second oscillation cycle of the carrier field (Fig. 3a), we track the spectrogram of a single Dirac electron (Fig. 3b) with a minimal quantum model (see Methods). The driving MIR field periodically modulates the phase of the electron wave function $\Psi(t)$. After one oscillation cycle, constructive superposition of amplitudes with the same phase becomes possible and leads to the appearance of peaks in the spectral density $|\Psi(\varepsilon)|^2$ located at the energies $\pm\hbar\Omega_{MIR}$. In the absence of dephasing, these Floquet sidebands continuously sharpen as the number of cycles increases. Sidebands of order $n$ gain relevant weight only if the driving field accelerates electrons to a quiver energy (Fig. 3b, black curve) of $n \times \hbar\Omega_{MIR}$ (see Methods).

For a more detailed microscopic description, we perform full quantum simulations of the subcycle ARPES spectra with MIR driving based on a nonequilibrium Green's functions approach (see Methods). The resulting spectra (Fig. 3c and Extended Data Fig. 9) reproduce the salient experimental features very well: starting from the ground-state distribution, the electrons are accelerated, initially within the original Bloch band. At early times, quantum interference is not strong enough to cause band splitting but becomes already visible in terms of an apparent steepening of the band ($t$ = -60 fs, Fig. 3c). Clearly discernible sideband splitting sets in near the second MIR field crest ($t$ = -50 fs) with the intensity distributed nearly equally between both branches. A quarter of an optical cycle later, the split intensity has shifted almost completely to the left ($t$ = -40 fs). A further half cycle later ($t$ = -20 fs), when most of the electrons have been accelerated to the right branch, also the MIR field has increased and most of the intensity now appears in the first Floquet-Bloch sideband.

Based on these calculations one can consistently understand both, intraband currents and the emergence of Floquet-Bloch bands, from the micromotion of electrons subject to a strong driving field.

Essentially, the Floquet states are the result of a rapidly growing phase space for quantum trajectories upon strong driving and their constructive and destructive interference. Importantly, the averaged electron momenta follow the classical expectation over many cycles although the trajectory of a single electron is well defined only in the subcycle regime (Fig. 3b, black curve). While the electrons move through the band structure they collect quantum phases, which define the shape and the weight of the Floquet-Bloch sidebands. The nontrivial nature of this process is clearly visible, e.g., from the fact that at $t = -20$ fs (Fig. 3c), the strongest weight does not reside in the equilibrium band as expected from the single-state picture of Fig. 3b, but is rather transferred almost entirely to the first sideband.

## A Floquet path for interband transitions

Our experiments also shed light on a long-standing question in HHG, namely, how multi-photon excitation between equilibrium bands proceeds in the strong-field limit. In the three-dimensional topological insulator $Bi_2Te_3$, the direct bulk bandgap at the $\Gamma$ point exceeds 0.3 eV and thus prevents resonant one-photon interband transitions[28] ($\hbar\Omega_{MIR} = 0.1$ eV). A finite density of electrons can nonetheless dynamically bridge the gap following an intriguing quantum path through Floquet-Bloch replicas of the TSS. This is best seen in subcycle ARPES maps (Fig. 4a) taken near and after the field crest of the MIR waveform. At the peak field of 0.8 MV cm$^{-1}$ (Fig. 4b), the electronic distribution deviates dramatically from the original band dispersion (Fig. 4a, $t = -12$ fs): a large fraction of the photoelectron signal is transferred to the first and second Floquet-Bloch sidebands, where the electrons are periodically slushed back and forth by the carrier field (Fig. 4a, $t = 22$ fs and 46 fs). The electron transfer changes critically on a few-fs scale. Since the experiment is performed under the conditions of HHG[28], our data capture electron dynamics underlying HHG directly in momentum space, for the first time. Once the high-energy tail of the Floquet electron distribution reaches the bulk conduction band (Fig. 4a, grey area) the surface states hybridize with the bulk band and populate it. The corresponding photoelectron signature persists even after the MIR pulse is over (Fig. 4c), confirming transitions into the bulk conduction band.

The novel mechanism of transitions between equilibrium bands via Floquet-Bloch states is faithfully reproduced by quantum-nonequilibrium calculations. Figure 4d shows simulated subcycle-ARPES spectra at $t$ = 22 fs and 46 fs. As in the experiment (Fig. 4a), strong photoelectron intensities are generated in an energy-momentum region where the second and third Floquet-Bloch sidebands overlap with the bulk states, creating a persistent bulk conduction band population. For a more quantitative analysis, we trace the energy-momentum integral of the measured photoelectron intensity over the bulk conduction band region (Fig. 4c, dashed trapezoid), $I_{BCB}$, as a function of $t$ (Fig. 4e). After a steep increase during the MIR pulse, $I_{BCB}$ settles to a long-lived plateau, reflecting the incoherent electron population scattered into the bulk conduction band. Remarkably, the onset of $I_{BCB}$ is superimposed with pronounced oscillations (Fig. 4e), which are fingerprints of the coherent coupling of the transiently dressed TSS with the bulk conduction band as qualitatively confirmed by calculations (Fig. 4f). The population of the Floquet surface states (red solid curve) strongly oscillates owing to inter- and intra-sideband excitations. When the dressed states reach the conduction band while the surface-bulk coupling is still coherent, the electron population can be transferred into and out of the bulk (grey curve) as observed experimentally. Ultimately, scattering creates an incoherent bulk population, which depletes the carrier density in the surface state as compared to a situation without bulk-surface coupling (red dashed curve) and accelerates the collapse of the Floquet band structure while the MIR pulse ends. The transition thus results from a highly nonlinear strong-field process of a class that has now become directly observable for the first time.

The ultrafast buildup and collapse of the dressed band structure open an unprecedented dynamical view of one of the essential conditions for Floquet-Bloch states: dephasing. For dressed states to form, electrons need to coherently follow the driving field and enable multi-path quantum interference. From the surprisingly fast emergence of Floquet sidebands after only one oscillation cycle (Fig. 2c), we conclude that the quantum memory must survive at least one MIR oscillation period. This condition is exquisitely fulfilled for the TSS, where relaxation times exceeding 1 ps have been observed[24]. Yet the bulk of $Bi_2Te_3$ forms a trivial insulator with relaxation times substantially shorter than $2\pi/\Omega_{MIR}$ (ref. [28]). This explains why we observe Floquet-Bloch sidebands of the TSS, while there is no trace of analogous replicas of the bulk valence band in the experimental ARPES maps (Figs. 2 and 4). Intriguingly, the electron transfer within the Floquet-Bloch bands dynamically changes coherence. When the dressed surface states can scatter into the bulk, they rapidly lose coherence, sealing the end of the dressed states. By increasing the driving frequency, we can boost this effect (Extended Data Fig. 10).

**Conclusions and outlook**

In conclusion, we truly visualized the birth, rise, and collapse of Floquet-Bloch states in a topological insulator, in both experiment and quantum theory. By introducing subcycle ARPES in the MIR, we directly visualize in momentum space how the lightwave-driven surface state of $Bi_2Te_3$ crosses over from an initial regime of deterministic quantum trajectories to a Floquet band structure resulting from multi-path quantum interference, within an Ehrenfest time[39] of one optical cycle. The dynamics unify the particle aspect of electrons describing subcycle acceleration with the wave-like interference of electrons forming light-induced sidebands. Ultimately, electrons promoted in the emerging Floquet-Bloch band structure can scatter into the bulk conduction band, destroying the Floquet states. Our capability to image band structures that are transiently dressed and accelerated by strong light fields marks a milestone towards a complete understanding of Floquet engineering, HHG, and other strong-field light-matter interaction concepts[40]. Also, Floquet topological insulator phases in graphene[1,9,14], light-induced topological phase transitions in Weyl semimetals[3,10], and dynamical Bloch oscillations triggering HHG bursts[18-28] may now be visualized on the subcycle scale and, potentially, reveal new ways to design quantum functionalities with light.

1. Oka, T. & Aoki, T. Photovoltaic Hall effect in graphene. *Phys. Rev. B* **79**, 081406 (2009).

2. Mitrano, M. *et al*. Possible light-induced superconductivity in $K_3C_{60}$ at high temperature. *Nature* **530**, 461-464 (2016).

3. Sie, E. J. *et al*. An ultrafast symmetry switch in a Weyl semimetal. *Nature* **565**, 61-66 (2019).

4. Basov, D. N., Averitt, R. D. & Hsieh, D. Towards properties on demand in quantum materials. *Nat. Mater.* **16**, 1077–1088 (2017).

5. de la Torre, A. *et al*. Colloquium: Nonthermal pathways to ultrafast control in quantum materials. *Rev. Mod. Phys.* **93**, 041002 (2021).

6. Lindner, N.H., Refael, G. & Galitski, V. Floquet topological insulator in semiconductor quantum wells. *Nat. Phys.* **7**, 490-495 (2011).

7. Wang, Y.H., Steinberg, H., Jarillo-Herrero, P. & Gedik, N. Observation of Floquet-Bloch States on the Surface of a Topological Insulator. *Science* **342**, 453-457 (2013).

8. Mahmood, F. *et al.* Selective scattering between Floquet-Bloch and Volkov states in a topological insulator. *Nat. Phys.* **12**, 306-310 (2016).

9. Sentef, M.A. *et al*. Theory of Floquet band formation and local pseudospin textures in pump-probe photoemission of graphene. *Nat. Commun.* **6**, 7047 (2015).

10. Hübener, H., Sentef, M.A., Giovannini, U.D., Kemper, A.F. & Rubio, A. Creating stable Floquet-Weyl semimetals by laser-driving of 3D Dirac materials. *Nat. Commun.* **8**, 13940 (2017).

11. Reutzel, M., Li, A., W, Z. & Petek, H. Coherent multidimensional photoelectron spectroscopy of ultrafast quasiparticle dressing by light. *Nat. Commun.* **11**, 2230 (2020).

12. Zhang, J. *et al.* Observation of a discrete time crystal. *Nature* **543**, 217-220 (2017).

13. Choi, S. *et al.* Observation of discrete time crystalline order in a disordered dipolar many-body system. *Nature* **543**, 221-225 (2017).

14. Mclver, J.W. *et al.* Light-induced anomalous Hall effect in graphene. *Nat. Phys.* **16**, 38-41 (2020).

15. Träger, N. *et al.* Real-Space Observation of Magnon Interaction with Driven Space-Time Crystals, *Phys. Rev. Lett.* **126**, 057201 (2021).

16. Ghimire, S. *et al.* Observation of high-order harmonic generation in a bulk crystal. *Nat. Phys.* **7**, 138-141 (2011).

17. Zaks, B., Liu, R. B. & Sherwin, M. S. Experimental observation of electron-hole recollisions. *Nature* **483**, 580-583 (2012).

18. Schubert, O. *et al.* Sub-cycle control of terahertz high-harmonic generation by dynamical Bloch oscillations. *Nat. Photon.* **8**, 119-123 (2014).

19. Luu, T. T. *et al.* Extreme ultraviolet high-harmonic spectroscopy of solids. *Nature* **521**, 498-502 (2015).

20. Langer, F. *et al*. Lightwave-driven quasiparticle collisions on a subcycle timescale. *Nature* **533**, 225-229 (2016).

21. Vampa, G. *et al.* Linking high harmonics from gases and solids. *Nature* **522**, 462-464 (2015).

22. Yoshikawa, N., Tamaya, T. & Tanaka, K. High-harmonic generation in graphene enhanced by elliptically polarized light excitation. *Science* **356**, 736-738 (2017).

23. Higuchi, T., Heide, C., Ullmann, K., Weber, H.B. & Hommelhoff, P. Light-field-driven currents in graphene. *Nature* **550,** 224-228 (2017).

24. Reimann, J. *et al*. Subcycle observation of lightwave-driven Dirac currents in a topological surface band. *Nature* **562**, 396-400 (2018).

25. Hafez, H. *et al.* Extremely efficient terahertz high-harmonic generation in graphene by hot Dirac fermions. *Nature* **561**, 507-511 (2018).

26. Langer, F. *et al.* Lightwave valleytronics in a monolayer of tungsten diselenide. *Nature* **557,** 76-80 (2018).

27. Borsch, M. *et al.* Super-resolution lightwave tomography of electronic bands in quantum materials. *Science* **370**, 1204-1207 (2020).

28. Schmid, C.P. *et al*. Tunable non-integer high-harmonic generation in a topological insulator. *Nature* **593**, 385-390 (2021).

29. Cheng, B. *et al*. Efficient Terahertz Harmonic Generation with Coherent Acceleration of Electrons in the Dirac Semimetal $Cd_3As_2$. *Phys. Rev. Lett.* **124**, 117402 (2020).

30. Hohenleutner, M. *et al.* Real-time observation of interfering crystal electrons in high-harmonic generation. *Nature* **523,** 572-575 (2015).

31. Schultze, M. *et al.* Attosecond band-gap dynamics in silicon. *Science* **346,** 1348-1352 (2014).

32. Hasan, M.Z. & Kane, C.L. Colloquium: Topological insulators. *Rev. Mod. Phys.* **82,** 3045 (2010).

33. Chen, Y.L. *et al.* Experimental Realization of a Three-Dimensional Topological Insulator, $Bi_2Te_3$. *Science* **325**, 178-181 (2009).

34. Holthaus, M. Floquet engineering with quasienergy bands of periodically driven optical lattices. *J. Phys. B: At. Mol. Opt. Phys.* **49**, 013001 (2016).

35. Ikeda, T.N., Tanaka, S. & Kayanuma, Y. Floquet-Landau-Zener interferometry: Usefulness of the Floquet theory in pulse-laser-driven systems, *Phys. Rev. Research* **4**, 033075 (2022).

36. Aeschlimann, S. *et al.* Survival of Floquet-Bloch States in the Presence of Scattering. *Nano Lett.* **21**, 5028-5035 (2021).

37. Zhang, P. *et al.* Observation of topological superconductivity on the surface of an iron-based superconductor, *Science* **360**, 182-186 (2018).

38. Miaja-Avila, L. *et al.* Laser-Assisted Photoelectric Effect from Surfaces. *Phys. Rev. Lett.* **97**, 113604 (2006).

39. Maldacena, J. *et al*. A bound on chaos. *J. High Energ. Phys.* **2016**, 106 (2016).

40. Ikeda, T.N. & Sato, M. General description for nonequilibrium steady states in periodically driven dissipative quantum systems, *Science Advances* **6**, eabb4019 (2020).

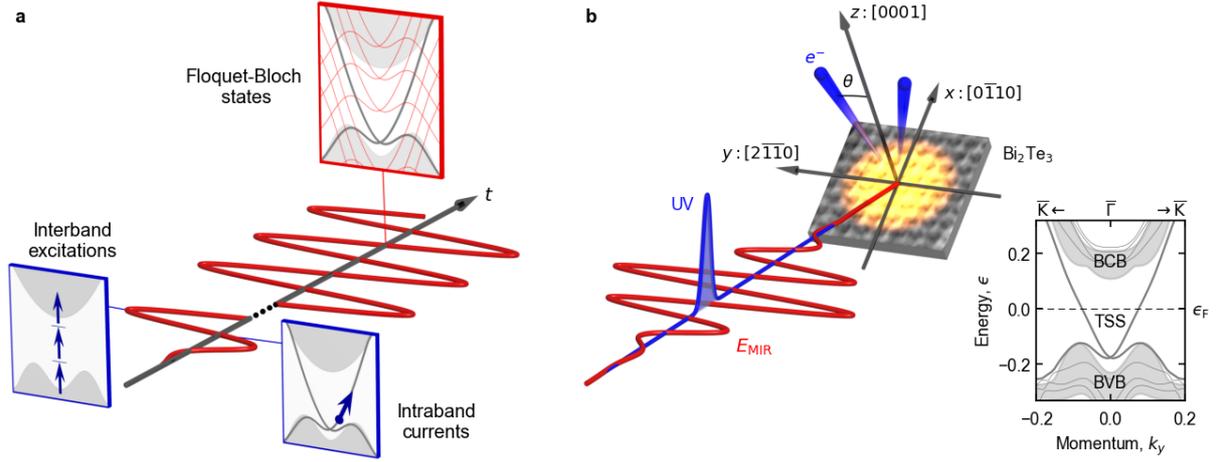

**Fig. 1 | Conceptual idea of Floquet band engineering on the subcycle scale. a**, Illustration of fundamental electronic processes induced by intense MIR light fields in solids. Whereas multiphoton interband excitation as well as lightwave-induced intraband currents are expected to occur already in the subcycle regime, strong-field periodic driving may result in Floquet-Bloch states. Yet the transition between these regimes is unclear. **b**, Schematic of the experiment: an s-polarized, phase-stable MIR electric field $E_{MIR}$ (red waveform) excites the topological surface state (TSS) of $Bi_2Te_3$. Angle-resolved photoelectron spectra recorded with p-polarized ultraviolet (UV) probe pulses (blue) at variable time delay provide snapshots of the induced dynamics in the electronic band structure $\epsilon(k_y)$ with subcycle temporal resolution. The direct mapping of the electronic band structure includes the bulk valence band (BVB) and the bulk conduction band (BCB).

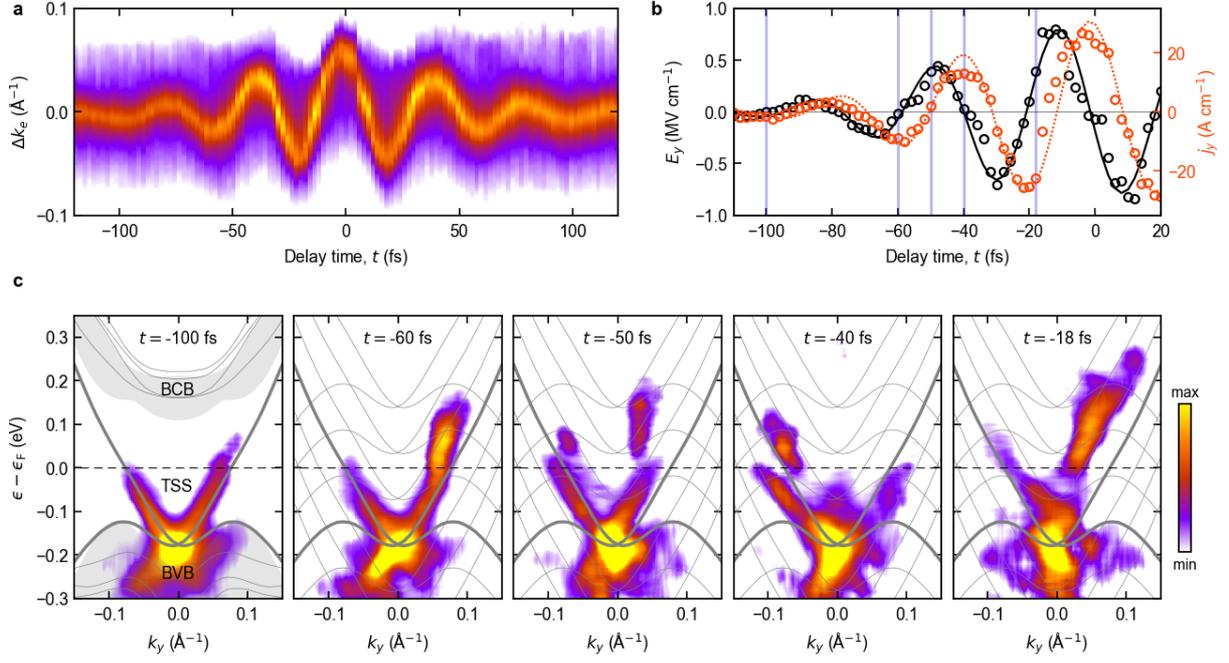

**Fig. 2 | Subcycle observation of Floquet-Bloch sidebands. a**, Curvature-filtered photoelectron momentum-streaking trace of the 25 THz driving field. **b**, First half of the electric-field waveform $E_y$ reconstructed from **a** (black circles) and current density $j_y$ extracted from the measured asymmetry in the population of the surface bands (red circles). The black solid curve is a fitted analytic function of the electric field with a peak amplitude of 0.8 MV cm⁻¹, the solid red line is the electric current induced by this waveform according to the semiclassical Boltzmann model. Thin vertical lines indicate temporal positions of the ARPES maps displayed in **c**. **c**, Curvature-filtered lightwave ARPES maps recorded before the arrival of the MIR field ($t$ = -100 fs) and at selected delay times during the first half of the driving pulse, at $t$ = -60 fs, -50 fs, -40 fs, and -18 fs. The images are compensated for the streaking shifts in **a**. The DFT band structure is superimposed in the leftmost panel. The electronic distributions are initially accelerated along the ground-state band dispersion, but subsequently divert from it, splitting into multiple branches that follow approximately the TSS band structure shifted by integer multiples of the driving frequency (thin grey curves).

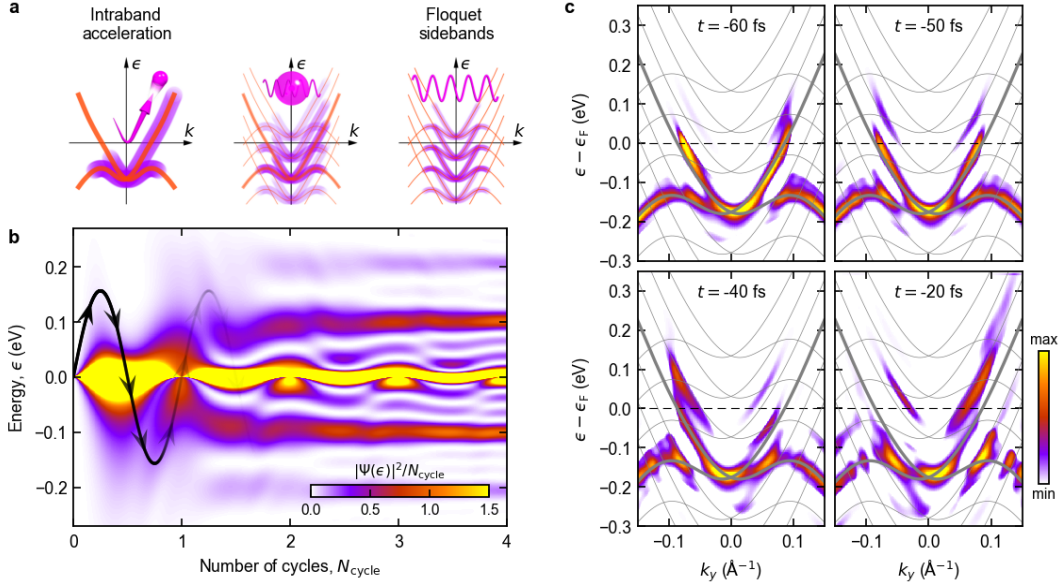

**Fig. 3 | Lightwave ARPES calculations capturing the birth of Floquet-Bloch states. a**, Illustration of light-matter interaction in the strong field regime: electron acceleration along a dispersing band as observed by a short probe pulse (left). Floquet-Bloch states observed by continuous-wave probing (right). Under the present experimental conditions, both aspects simultaneously modulate the electronic distribution on the subcycle time scale (centre). **b**, Dependence of Floquet-Bloch energy spectra on the number of pump optical cycles, obtained from a minimal quantum model (see Methods). The black curve indicates the trajectory of a massless Dirac electron (Fermi velocity $v_F = 4.1$ Å/fs) in the periodic driving field. **c**, Lightwave ARPES maps calculated at four temporal steps of $t$ = -60 fs, -50 fs, -40 fs, and -20 fs. Curvature image filtering has been applied in consistency with the experimental data in Fig. 2c. See Methods for the further details of the filtering procedure. Grey curves indicate the TSS band structure as derived from the $\boldsymbol{k} \cdot \boldsymbol{p}$ Hamiltonian (thick lines) and shifted by integer multiples of the driving frequency (thin lines).

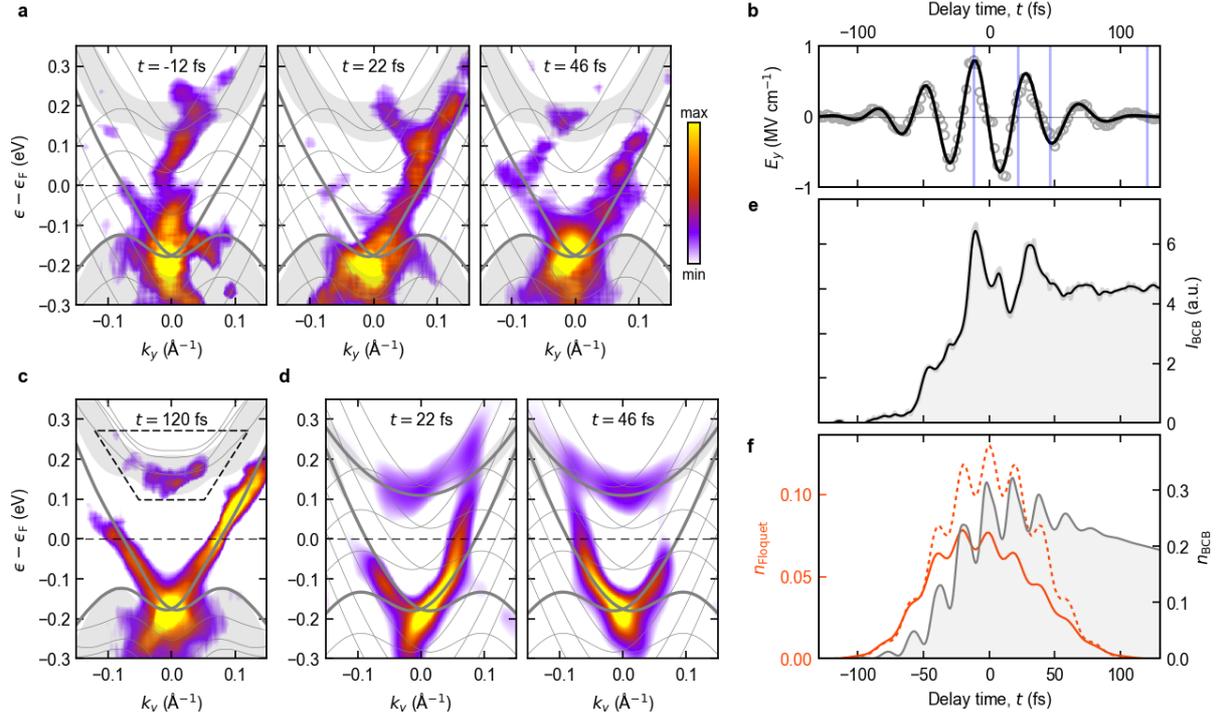

**Fig. 4 | Interband transition to the bulk conduction band via dynamical Floquet-bulk coupling. a**, Curvature-filtered lightwave ARPES maps as in Fig. 2c but measured at $t$ = -12 fs, 22 fs, and 46 fs. **b**, Experimental MIR driving waveform. Grey markers show the electric field from the streaking reconstruction, the black solid curve indicates its fit as in Fig. 2b. Thin vertical lines highlight the temporal steps in **a**. **c**, Measured ARPES maps after MIR excitation ($t$ = 120 fs). **d**, Curvature-filtered ARPES maps calculated at $t$ = 22 fs and 46 fs with coupling of the TSS to bulk states. Thick grey curves describe the model band structure as in Fig. 3c. Grey shaded areas illustrate delocalization of the bulk states to account for self-energy effects in our model (see Methods). **e**, Temporal evolution of the experimental photoemission intensity from the bulk conduction band states obtained by integrating the raw ARPES data over the trapezoid shown in **c**. Grey curve and shade, experimental data; black curve, smoothed guide to the eye. **f**, Rise and decay of the population of Floquet surface states $n_{Floquet}$ obtained from the calculated photoemission intensity with (red solid line) and without (red dashed line) coupling to bulk states. The grey curve describes the temporal evolution of the calculated occupation of the bulk conduction band $n_{bulk}$.

**Methods**

**Experimental setup.** The optical part of the lightwave ARPES setup (Extended Data Fig. 1a) is based on a titanium-sapphire (Ti:Sa) laser amplifier generating near-infrared light pulses with a duration of 33 fs, a centre wavelength of 807 nm, and a pulse energy of 5.5 mJ, at a repetition rate of 3 kHz. A first branch of the laser output pumps two parallel dual-stage optical parametric amplifiers (OPAs)[41], whose signal wavelengths are set to 1.2 μm and 1.3 μm, respectively. Because both OPAs are seeded by the same white-light continuum they share the same CEP fluctuations. Difference frequency mixing between the two signal pulse trains in a gallium selenide crystal (thickness, 430 μm), thus, generates passively phase-locked MIR pulses. Their CEP can be set by the relative time delay between the two near-infrared pulses, $t_{CEP}$, whereas the MIR centre frequency can be tuned from 25 THz to 41 THz by varying the signal wavelengths of the OPAs and adapting the phase-matching angle of the mixing crystal. A 500-μm thick Brewster window of germanium separates the residual near-infrared light from the MIR beam. The polarization and the power of the MIR pulse are precisely controlled by a pair of wire-grid polarizers (WG). A typical MIR waveform detected by electro-optic sampling in a 6.5-μm zinc telluride crystal is displayed in Extended Data Fig. 1b; the inset shows its amplitude spectrum.

A second branch of the laser fundamental is focused onto four thin quartz plates under Brewster's angle to generate a broad multi-plate continuum (MPC)[42]. The input pulse energy is adjusted by a half-wave plate ($\lambda/2$) and a thin-film polarizer (P), and the focusing condition is fine-tuned via a lens (L) and an adjustable iris aperture (A1). The laser fundamental and the resulting MPC spectra are shown in Extended Data Fig. 1c. The generated MPC pulse is compressed to 14 fs, by a first set of prisms (P1, P2), as confirmed by frequency-resolved optical gating (FROG). The compressed MPC pulse is then focused onto a 30-μm β-BBO crystal to generate its second harmonic, which is subsequently compressed by a second pair of prisms (P3, P4). Its intensity spectrum is centred at a wavelength of 410 nm and features a bandwidth of 25 nm (Extended Data Fig. 1d). The pulse duration is measured by cross-correlation FROG (XFROG) with a gate pulse derived from a part of the MPC light that is passed through a 10-nm bandpass filter to obtain a well-defined temporal profile with a duration of 100 fs. Extended Data Fig. 1e compares measured and reconstructed XFROG traces, and the intensity and phase profiles of the reconstructed pulse are shown in Extended Data Fig. 1f with a full width at half maximum (FWHM) of 17 fs.

The intense MIR pump and the broadband UV probe are spatially overlapped with a thin indium tin oxide (ITO) window and coupled into the ultrahigh vacuum (UHV) chamber through a 1-mm diamond window. The UHV chamber is shielded by a mu-metal liner and maintains a base pressure of $3 \times 10^{-11}$ mbar. The spatially overlapped beams are focused onto the sample under an angle of incidence of 53° by a UV-grade aluminium focusing mirror (focal length, 75 mm) on a UHV-compatible, piezo-controlled mirror mount inside the chamber. Photoelectrons are emitted via a two-photon photoemission (2PPE) process of the UV probe pulse, whose total excitation energy corresponds to 6.0 eV. The emitted electrons are detected by a hemispherical electron analyser with an image detector (Specs Phoibos 150). All measurements are performed at room temperature. The energy resolution is governed by the broad spectrum of the probe pulse, which can introduce an energy broadening of $100 \pm 10$ meV as determined from a band splitting in raw ARPES spectra (Extended Data Fig. 5). The pulse compression on the sample surface was optimized by maximizing the 2PPE counts with the position of prism P4 (Extended Data Fig. 1g). For the XFROG analysis to be applicable to the actual measurement condition, exactly the same optical components (ITO and diamond windows) were inserted in the XFROG setup.

**Sample preparation.** The high-quality $Bi_2Te_3$ sample was grown by using the modified Bridgman technique[43]. A gradual modulation in the carrier concentration was created by an appropriate solidification condition, and a well-defined p-type sample was chosen for the present experiment. Transport measurements revealed a high carrier mobility of $10^4$ cm$^2$ V$^{-1}$ s$^{-1}$ at a temperature of 4 K. An atomically flat surface was obtained by cleaving the sample *in-situ* in the UHV chamber typically at $2 \times 10^{-10}$ mbar. The Dirac point of the sample is located 0.18 eV below the Fermi level.

**DFT calculations of the band structure of $Bi_2Te_3$.** The ground-state $Bi_2Te_3$ band structure was obtained from density-functional theory (DFT) calculations performed with the QUANTUM ESPRESSO code[44]. The generalized gradient approximation was chosen for the exchange-correlation functional, and an ultrasoft pseudopotential with full-relativistic spin-orbit coupling was used. A freestanding $Bi_2Te_3$ slab with a thickness of 5 quintuple layers (QLs) and a vacuum region of a length of 1 nm were included. The lattice parameters of the slab were fixed to the experimental values of ref.[45]. A $10 \times 10 \times 1$ grid of k-point sampling was used. Convergences with respect to the energy cut-off, the k-point sampling, and the vacuum length were confirmed.

**Curvature-based image processing for enhanced band visibility.** The probe spectrum supporting subcycle resolution inevitably broadens the measured ARPES spectra. Extended Data Figs. 2a and b show ARPES maps measured by using a narrow-band 200-nm probe and the broadband 2PPE probe, respectively. The former shows a sharp Dirac dispersion in excellent agreement with the DFT calculations (grey curves). Whereas the latter still exhibits two Dirac branches following the grey curves, it suffers from significant spectral broadening and strong intensity from the bulk valence band. The dominating strength of the bulk signal is likely caused by a photoemission matrix element unique to the 2PPE probe condition. A well-established procedure to enhance the visibility of photoemission and other spectra is to plot the second derivative of the measured data. One known drawback is that the second derivative can introduce small shifts of peak positions. This problem can be overcome by a more sophisticated approach: a curvature method[37,46]. The second derivative of a function is connected to its curvature by

$$C(x) = \frac{f''(x)}{(C_0 + f'(x)^2)^{3/2}} \; . \tag{1}$$

Here, the factor $C_0$ controls precision and sharpness of extracted peaks: in the limit of $C_0 \rightarrow \infty$, the formula is reduced to the second derivative $f''(x)$, whereas for $C_0 \rightarrow 0$ the formula is approximated by $\frac{f''(x)}{f'(x)^3}$, which diverges when $f'(x) = 0$, that is, at peak positions. The one-dimensional method can be naturally extended to the two-dimensional (2D) case of ARPES maps as introduced in ref.[46]. It has been shown that $C_0$ is conveniently given in units of the maximum $f'(x)$ value; we consistently use $C_0 = 0.01$. Reference[46] also demonstrated that it is possible to tune the peak visibility by a smoothing process applied prior to curvature image processing. This facilitates the quantitative extraction of lightwave-induced currents from ARPES maps by intentionally blurring Floquet fine structures (Extended Data Fig. 3). For illustration, Extended Data Fig. 2c shows the broadband ARPES map processed by four linear convolutions with a smoothing box size of 0.035 Å$^{-1}$ × 0.06 eV and subsequent curvature image filtering. Specifically, we utilize the 2D curvature filtering described by Eq. (11) of ref.[46]. To eliminate lattice-like artefact patterns generated during the numerical derivation, we also apply a 2D median filter within a box size of 0.035 Å$^{-1}$ × 0.06 eV. The procedure significantly enhances the visibility of the two Dirac branches following the DFT band structure and suppresses the strong bulk intensity. Using a larger smoothing box of 0.05 Å$^{-1}$ × 0.09 eV with the same factor blurs the filtered image accordingly (Extended Data Fig. 2d). It facilitates detecting coarse structures and quantifying the population of both branches of the TSS.

As is the case for any type of image filters, the curvature filtering could, in principle, introduce artificial features. We therefore carefully confirmed that all relevant spectral features such as the sideband signatures are also present in the raw ARPES images as shown in Extended Figs. 5 and 6.

**Electric field reconstruction from vacuum streaking.** Vacuum streaking of photoelectrons allows for *in-situ* reconstruction of electric field waveforms from lightwave ARPES data. The procedure is particularly straightforward for s-polarized pump electric fields because they are continuous at the surface. Thus, the momentum change of the emitted photoelectrons equals the integral of the force exerted on the electrons. The electric field is then simply given by the time derivative of the momentum streaking trace $k_{\text{streak}}(t)$ (ref.[24]). Extended Data Fig. 3a shows a momentum-streaking trace obtained from a 25 THz pump field. The intensity profile at each time step is obtained by integrating raw ARPES spectra around the strong bulk peak. Curvature filtering enhances the visibility of the streaking traces (Extended Data Fig. 3b). Extended Data Fig 3c displays the extracted streaking waveform $k_{\text{streak}}(t)$ and the surface electric field $E_y(t)$. The waveform is well described by a sinusoidal curve with a Gaussian envelope,

$$E_y(t) = E_0 e^{-\frac{t^2}{\tau^2}} \sin(\Omega_{\text{MIR}} t + \varphi_{\text{CEP}}).  \qquad (2)$$

With a value of 17 fs, the duration of the probe pulse is shorter than the MIR oscillation period (40 fs at 25 THz) but not short enough to neglect it for determining the amplitude of the streaking waveform. A linear convolution of a 25-THz input waveform and a 17-fs Gaussian pulse would lead to a reduction of the measured amplitude by ~0.5. Owing to the nonlinearity of the 2PPE processes the actual scaling factor is closer to one. By means of lightwave ARPES calculations, we determined the factor to be 0.8 for the 25 THz waveform. At a MIR frequency of 41 THz, the factor reduces to 0.4. All waveform amplitudes given in the present work have been corrected by the appropriate scaling factors. The field strength determined at the surface is reduced with respect to the incoming field strength due to screening by the sample. This estimated reduction factor is 9 in the case of $Bi_2Te_3$[28] and under our experimental geometry.

**Streaking compensation of ARPES data.** All ARPES maps have been compensated for streaking effects using the procedure described in ref.[24]. In doing so, we neglect the small energy shift that the photoemitted electrons gain in the s-polarized pump electric field. For weak momentum streaking $\Delta p = \hbar \Delta k$, the shift $\Delta \varepsilon$ is proportional to initial parallel velocity $\mathbf{v}_{0,||}$ of the photoelectrons ($\Delta \varepsilon = \Delta \mathbf{p} \cdot \mathbf{v}_{0,||}$) and thus slightly tilts the ARPES maps of the linearly dispersing Dirac bands[24]. For the maximum observed streaking of $\Delta k = 0.07$ Å$^{-1}$, the resulting change of the apparent Fermi velocity is $\Delta v_F = (\hbar/m)\Delta k = 0.81$ Å fs$^{-1}$, i.e., it amounts to 20% the values of $v_F$ measured without a pump field.

**Current reconstruction.** The driving MIR field induces an asymmetry of the electron population between the left and right branch of the TSS. This asymmetry corresponds to a surface current. Extended Data Fig. 3d shows lightwave ARPES maps compensated for the streaking shifts in Extended Data Fig. 3a at three temporal positions of $t$ = -100 fs, 0 fs, and 20 fs. The latter two correspond to time steps with maximum excursions in the left and right Dirac branches, respectively. The driving MIR field with an amplitude of up to 0.5 MV cm$^{-1}$ is already high enough for the detection of Floquet sidebands (compare Extended Data Fig. 4a). These sidebands have been averaged out and suppressed in the ARPES maps of Extended Data Fig. 3a by appropriate tuning of the curvature filtering (Extended Data Fig. 2d). The suppression facilitates the reliable analysis of oscillating currents in the presence of Floquet sidebands. Following the approach of ref.[24], spectral intensities are integrated separately for the left and right branches in the blue and red boxes and converted to the current waveform plotted in the bottom panel of Extended Data Fig. 3e (red circles). For comparison, the solid curve indicates the current predicted by a scattering-free Boltzmann equation using the measured electric field (top panel). The excellent quantitative consistency demonstrates not only the robustness of the semiclassical acceleration picture even under the conditions of a strong and rapidly oscillating MIR field causing huge electron redistributions in the Dirac cone. It also confirms that curvature filtering, if carefully applied, allows for a quantitative evaluation of the lightwave ARPES data.

**Observation of Floquet-Volkov sidebands with p-polarized pump.** The interaction of photoelectrons with the periodic electric field of the pump pulse can lead to Volkov sidebands with the same amount of splitting as Floquet-Bloch states[8,36,38,47]. This so-called laser-assisted photoemission effect is known to be strongly suppressed under the conditions of our experiment when the purely s-polarized pump electric field lies in the surface plane[8]. The calculations discussed below predict that 80-90% of the observed sideband intensity is composed of pure Floquet states and only 10-20% stems from Volkov states. As an experimental test, we measured lightwave ARPES maps with mixed s- and p-polarized MIR pump fields (Extended Data Fig. 7). Although the overall MIR field strength is lower, the sideband signatures are stronger than in Figs. 2 and 4. This result is fully consistent with ref. [8], where multiple sidebands for Volkov states and only the first sideband for pure Floquet states were observed. Even though the split-off bands reach up to the position of the bulk conduction band, no population is left at later times (Extended Data Fig. 7c, $t = 90$ fs). This apparent lack of coupling and the notably different fine structure of the split bands of Fig. 2 and Extended Data Fig. 7 clearly point to the distinct origins of the Floquet and Volkov states. For pure in-plane pumping with s-polarized MIR light, the fine structure of the recorded ARPES maps thus originates from real photon-dressing in matter, i.e., Floquet-Bloch states, and not from LAPE/Volkov contributions.

**Absence of 2PPE resonances in Bi$_2$Te$_3$.** Unoccupied surface states of Bi$_2$Te$_3$ that are resonantly excited with the 400-nm probe pulses could potentially modulate the ARPES intensity distribution measured by a 2PPE process. In extreme cases, this could even introduce artificial sideband structures, completely unrelated to Floquet-Bloch states. The same holds for final-state resonances above the vacuum level. To exclude this possibility, we performed systematic 2PPE measurements with a picosecond, wavelength-tuneable light source. Extended Data Fig. 8a shows the corresponding spectra (blue curves) together with the broadband UV spectrum used in lightwave ARPES (orange curve). The centre wavelength of the light source is widely tuned covering the full range of the broadband spectrum while the intensity is controlled such that it introduces a constant number of photons. 2PPE-ARPES maps measured by the tuneable probe with six different centre wavelengths are plotted in Extended Data Fig. 8b. The almost identical intensity distribution clearly excludes that any of the sidebands observed upon MIR pumping are artefacts of the 2PPE probe process. Moreover, we find that matrix element effects of the measured ARPES intensity distribution are generally small over the entire energy range (Extended Data Fig. 8c).

**Minimal model of Floquet-Bloch states.** A simple model of Floquet-Bloch states is based on a two-dimensional massless Dirac Hamiltonian

$$H_0(\boldsymbol{k}) = \hbar v_F (\sigma_x k_y - \sigma_y k_x),\qquad(3)$$

subject to the light-matter interaction in the dipole gauge: $\hbar\boldsymbol{k} \to \hbar\boldsymbol{k} - q\boldsymbol{A}(t)$ (ref. [48]). $\sigma_i$ ($i = x, y$) denote Pauli matrices, $v_F$ the Fermi velocity, $q$ the electron charge, and $\boldsymbol{A}(t)$ the vector potential of the driving field $\boldsymbol{E}(t) = -\partial\boldsymbol{A}(t)/\partial t$ . Since we apply a linearly polarized pump electric field in the y-direction and focus on dynamics along the same direction, the time-dependent Hamiltonian becomes

$$H(k_y, t) = (\hbar v_F k_y - q v_F A_y(t))\sigma_x.\qquad(4)$$

It can be readily diagonalized, and the solution of the Schrödinger equation $i\hbar(\partial/\partial t)|\Psi(k_y,t)\rangle = H(k_y,t)|\Psi(k_y,t)\rangle$ is given by

$$|\Psi_\alpha(k_y,t)\rangle = \exp\left[-\frac{i}{\hbar}\left\{\alpha\hbar v_F k_y t - \alpha q v_F \int_{-\infty}^{t} dt'\, A_y(t')\right\}\right]|\psi_\alpha\rangle,\qquad(5)$$

where $\alpha$ denotes the two Dirac branches ($\alpha = \pm 1$) and $|\psi_\alpha\rangle$ are eigenstates of the ground-state Hamiltonian. In the case of continuous-wave driving, $A_y(t) = A_0 \sin \Omega t$, the wave function can be further transformed to

$$|\Psi_\alpha(k_y,t)\rangle = \exp\left\{-\frac{i}{\hbar}\left(\alpha\hbar v_F k_y t + \alpha q v_F \frac{A_0}{\Omega}\cos\Omega t\right)\right\}|\bar\psi_\alpha\rangle$$

$$= \left[\sum_{m=-\infty}^{\infty} (-i)^m J_m\left(\frac{\alpha q v_F A_0}{\hbar\Omega}\right)\exp\left\{-\frac{i}{\hbar}(\alpha\hbar v_F k_y + m\hbar\Omega)t\right\}\right]|\bar\psi_\alpha\rangle.\qquad(6)$$

Here, the Fourier expansion $e^{-ix\cos\Omega t} = \sum_{m=-\infty}^{\infty} (-i)^m J_m(x) e^{-im\Omega t}$ is used, where $J_m$ is the $m$-th Bessel function of the first kind, a constant phase factor is included to $|\tilde{\psi}_\alpha\rangle$, and $A_0 = E_0/\Omega$. In this analytical solution, one clearly recognizes Floquet-Bloch states with quasi-energies $\alpha\hbar v_F k_y + m\hbar\Omega$ and weights $J_m(\alpha q v_F E_0/\hbar\Omega^2)$. Although Floquet-Bloch states can be defined even with weak electric fields, they have finite amplitudes only with sufficiently large $E_0$ in relation to the square of the driving frequency $\Omega$. They are thus a hallmark of the strong-field regime. For $E_0 = 0.6$ MV cm$^{-1}$ and $\Omega = 25$ THz, as in our experiment, the argument of $J_m$ is $q v_F E_0/\hbar\Omega^2 = 1.51$ and we obtain $J_0 = 0.50$, $J_1 = 0.56$, i.e., a similar amplitude of ground state and first sideband in the Floquet limit. The relative amplitude of higher sidebands decreases rapidly with increasing $m$ ($J_2 = 0.24$, $J_3 = 0.06$, $J_4 = 0.01$).

For finite duration of the driving field, the spectral density $|\Psi_\alpha(k_y, \omega)|^2$ can be evaluated numerically by Fourier transforming Eq. (5). By performing this calculation with a continuously increasing time duration of the sinusoidal driving field, we obtain the Floquet spectrogram Fig. 3b. This analysis allows us to track the temporal evolution of Floquet-Bloch states independently of probing conditions. It clearly shows that the appearance of sidebands in the second optical cycle (Fig. 2) is an intrinsic property of the system. The sidebands evolve because the energy of each state at $k_y$ of the Dirac band oscillates as a function of time owing to strong driving by the electric field. Experimentally, we observe these oscillations additionally in terms of a redistribution of band population – that is, intraband currents – between the left and right branches of the Dirac surface states. Microscopically, the oscillations affect all electrons and are clearly visible as wiggling of the $E = 0$ band in Fig. 3b. In the absence of dephasing, the sidebands get sharper and thus appear more pronounced as the number of cycles increases. Figure 3b also shows that the onset of the time-dependent interaction leads to broad spectral wings of the initially sharp, unperturbed state, as the driven states are not energy eigenstates anymore. The spectral wings cover a wider energy range the stronger the driving field, before they gradually develop into clearly discernible sidebands. Higher order side bands evolve together with the first one with a relative intensity as for continuous wave driving. The spectral dispersion of the initially sharp eigenstate can be interpreted as a rapidly growing phase-space for quantum trajectories of a single electron. After one full oscillation cycle of the driving field, various trajectories then interfere constructively to form the emerging Floquet states spaced by the central photon energy.

We finally note that the maximum energy gain of a particle-like massless Dirac electron in a sinusoidal driving field is $\epsilon_{\max} = q v_{\mathrm{F}} A_0$. Under our experimental conditions, $\epsilon_{\max}$ exceeds the photon energy $\hbar\Omega$ of the driving field (black line in Fig. 3b). As Eq. (6) clearly shows, the larger the ratio $\epsilon_{\max}/\hbar\Omega$, the more important interference effects become for the temporal evolution of the system.

**Time-resolved ARPES calculations.** The study of the realistic behaviour of Floquet-Bloch states, such as dissipation, decoherence, and hybridization between multiple bands, calls for a more comprehensive theory beyond the minimal model. For this purpose, we introduce a general framework for simulating time-resolved ARPES spectra on a nonequilibrium Green's functions approach. The surface states of $\mathrm{Bi_2Te_3}$ are described by the $\boldsymbol{k} \cdot \boldsymbol{p}$ model from refs. [28,49], which is defined by the Hamiltonian

$$H_{\mathrm{TSS}}(\boldsymbol{k}) = C_0 + C(\boldsymbol{k})\boldsymbol{k}^2 + A(\sigma_x k_y - \sigma_y k_x) + 2R(\boldsymbol{k})(k_y{}^3 - 3k_x{}^2 k_y)\sigma_z. \tag{7}$$

The coefficients $A$, $C_0$, $C(\boldsymbol{k})$, and $R(\boldsymbol{k})$ for $\mathrm{Bi_2Te_3}$ are given in ref. [49], and the band dispersion is shown by red curves in Extended Data Fig. 9a (left). Here, the light-matter interaction is included in the velocity gauge[50] to facilitate calculations of time-resolved ARPES spectra as discussed below. Since the Hamiltonian (7) originates from $\boldsymbol{k} \cdot \boldsymbol{p}$ theory, the in-plane velocity matrix elements can be directly obtained from $\boldsymbol{v}(\boldsymbol{k}) = \nabla_{\boldsymbol{k}} H_{\mathrm{TSS}}(\boldsymbol{k})/\hbar$. The time-dependent Hamiltonian is thus given by

$$H(\boldsymbol{k}, t) = H_{\mathrm{TSS}}(\boldsymbol{k}) - q\boldsymbol{A}(t) \cdot \boldsymbol{v}(\boldsymbol{k}) + \frac{q^2 \boldsymbol{A}(t)^2}{2m}. \tag{8}$$

The dynamics of the Hamiltonian (8) is computed in the presence of dissipation and decoherence using the Liouville equation

$$\frac{d}{dt}\rho(\boldsymbol{k}, t) = -\frac{i}{\hbar}[H(\boldsymbol{k}, t), \rho(\boldsymbol{k}, t)] + D[\rho(\boldsymbol{k}, t)]. \tag{9}$$

We employ the relaxation-time approximation to the relaxation operator $D[\rho]$:

$$D[\rho(\boldsymbol{k},t)] = \begin{pmatrix} \dfrac{\rho_{00}(\boldsymbol{k},t) - n_0(\boldsymbol{k},t)}{T_1} & \dfrac{\rho_{01}(\boldsymbol{k},t)}{T_2} \\ \dfrac{\rho_{10}(\boldsymbol{k},t)}{T_2} & \dfrac{\rho_{11}(\boldsymbol{k},t) - n_1(\boldsymbol{k},t)}{T_1} \end{pmatrix}. \tag{10}$$

Here, $n_{0,1}(\boldsymbol{k},t)$ is the occupation in thermal equilibrium with respect to the instantaneous Hamiltonian $H(\boldsymbol{k},t)$ (instead of the equilibrium Hamiltonian). This procedure eliminates spurious behaviour in the standard relaxation-time approach in the presence of strong driving[51]. In our simulations we fix $T_2 = 10$ fs as the effective interband dephasing time[28,52], while for $T_1$ a large value of 1000 fs is chosen to reflect the long lifetime of the surface state protected by spin-momentum locking[24]. Calculations with a short $T_1$ value (1-10 fs) indeed show strong suppression of Floquet signatures in consistency with the latest study in graphene[36]. By systematically tuning $T_1$ and $T_2$ within a reasonable range, we have ensured that the actual choice of $T_{1,2}$ takes only a minimal effect on signatures of Floquet-Bloch states in the simulated time-resolved ARPES maps. The robustness is supported by the fact that electron-electron, electron-phonon, and other (e.g. disorder-mediated) scattering time scales are significantly longer than that of the observed photodressing. This simple fact manifests universality of our discovery: on sufficiently fast time scales, any light-matter-coupled system should exhibit the fundamental physics observed in the present, particularly clean testbed of the TSS.

After obtaining the time-dependent density matrix, we compute the lesser Green's function $G^<(\boldsymbol{k},t,t')$ as the central ingredient for the theory of time-resolved ARPES[53]. We employ the generalized Kadanoff-Baym ansatz (GKBA)[54] and obtain

$$G^<(\boldsymbol{k},t,t') = iU_{\boldsymbol{k}}(t,t')\rho(\boldsymbol{k},t'), t > t', \tag{11}$$

where $U_{\boldsymbol{k}}(t,t')$ is the time-evolution operator with respect to $H(\boldsymbol{k},t)$. Following ref.[55], the time-resolved ARPES signal is then computed from

$$I(\boldsymbol{k},\varDelta t) \propto \mathrm{Im} \operatorname{Tr} \sum_{\alpha} \int_0^{\infty} dt \int_0^{t} dt' s(t-\varDelta t) s(t'-\varDelta t) M_{\alpha}(\boldsymbol{k}) G^<_{\alpha\alpha}(\boldsymbol{k},t,t') M_{\alpha}^*(\boldsymbol{k}) e^{-i\varPhi(\boldsymbol{k},t,t')}, \tag{12}$$

where $s(t)$ denotes the envelope of the probe pulse, while

$$\Phi(\boldsymbol{k}, t, t') = \frac{1}{\hbar} \int_{t'}^{t} d\bar{t} \left[ \varepsilon_{\boldsymbol{p}}(\bar{t}) - \hbar\omega_{\mathrm{pr}} \right]. \tag{13}$$

The phase factor (13) accounts for the dressing of the photoelectron states $\varepsilon_{\boldsymbol{p}}(t) = (\boldsymbol{p} - q\boldsymbol{A}(t))^2/2m$; $\hbar\omega_{\mathrm{pr}}$ is the photon energy of the probe pulse. The in-plane component of the photoelectron momentum $\boldsymbol{p}$ is determined by momentum conservation: $\boldsymbol{p}_{\parallel} = \boldsymbol{k}$. The out-of-plane component is set by the photoelectron energy and the kinematics of the ARPES setup; we fix the value $p_{\perp} = 0.43 \text{ Å}^{-1}$ corresponding to the Dirac point. To understand the impact of LAPE we can also switch it off by neglecting the photodressing of the photoelectrons, which amounts to $\Phi(\boldsymbol{k}, t, t') \rightarrow (\varepsilon_{\boldsymbol{p}} - \hbar\omega_{\mathrm{pr}})(t - t')/\hbar$. We incorporated photoemission matrix elements $M_{\alpha}(\boldsymbol{k})$ in Eq. (12) (here $\alpha$ denotes the band index) and chose a simple model consistent with the experimental observation that the intensity is suppressed for larger $|\boldsymbol{k}|$:

$$M_{\alpha}(\boldsymbol{k}) = M_0 e^{-\frac{k^2}{2k_c^2}} \tag{14}$$

The cut-off momentum $k_c$ can be estimated from comparing spectra in equilibrium; we fix $k_c = 0.1 \text{ Å}^{-1}$. For simplicity, we assume the same dependence for both bands.

We have restricted the modelling of the time-resolved ARPES signal to single photon photoemission in order to limit the computational cost. In the experiment, however, the nonlinearity of the 2PPE probe results in a slightly higher spectral resolution than the bandwidth of the UV probe pulses. For the same reason, the effective time-resolution is in fact slightly better than the measured probe pulse duration. We consider this in the calculations by using a sech-shaped pulse form $s(t)$ which has, for the same pulse duration, a narrower bandwidth as compared to Gaussian pulses and matches the bandwidth to the experimentally observed resolution of the ARPES spectra. Such calculated raw and curvature filtered ARPES spectra for different delays are depicted in Extended Data Fig. 9c and d. The waveform shown in Extended Data Fig. 9b reproduces the experimental result in Fig. 2b. The chosen bandwidth of 47 meV (FWHM) allows for the clearest observation of both, the subcycle intraband acceleration and the formation of the Floquet sidebands, in excellent agreement with our experiment in Fig. 2. Extended Data Fig. 9e shows the impact of the bandwidth and the probe pulse form on the ARPES spectra. With decreasing bandwidth and therefore increasing pulse duration, the Floquet sidebands become more clearly visible, but the information on the subcycle dynamics gets lost. Although the experimental probe pulse might neither perfectly match a sech, a Lorentzian nor a Gaussian pulse, the comparison of the calculated 2PPE spectra for these three pulse forms show that the pulse form does not qualitatively change the results as long as the bandwidth is matched accordingly.

**Introduction of bulk states in the time-resolved ARPES calculations.** To investigate the effect of the unoccupied bulk states on the dynamics, we extend the Hamiltonian (7) by including additional bands. As a realistic model, we approximate the unoccupied bands with highest surface localization, as obtained from the DFT calculations performed on the 5-QL slab[56], by parabolic bands. The resulting model is referred to as a 3B model in what follows. The Hamiltonian for our 3B model reads

$$H_{3B}(\boldsymbol{k}) = \begin{pmatrix} H_{TSS,00}(\boldsymbol{k}) & H_{TSS,01}(\boldsymbol{k}) & 0 \\ H_{TSS,10}(\boldsymbol{k}) & H_{TSS,11}(\boldsymbol{k}) & 0 \\ 0 & 0 & \epsilon_R(\boldsymbol{k}) \end{pmatrix},$$
(15)

where $\epsilon_R(\boldsymbol{k}) = c_R + a_R \boldsymbol{k}^2$. The additional band $\epsilon_R(\boldsymbol{k})$ represents the surface-bulk resonance state. As shown by the DFT calculations, this band is located outside the bulk projected bands at the $\Gamma$ point, while becoming increasingly delocalized away from the Brillouin zone center. The quasi-surface-like character of the additional band is further justified by persistent time-resolved ARPES intensity after the pump pulse. The gradual transition to delocalized bulk states is incorporated by self-energy effects, as explained below. The band structure of the 3B Hamiltonian, including the width of the broadened bands in the region of the bulk projected bands, is shown in Extended Data Fig. 9a (right).

We incorporate light-matter interaction by including off-diagonal velocity matrix elements in addition to the intraband acceleration:

$$\boldsymbol{v}_{3B}(\boldsymbol{k}) = \frac{1}{\hbar} \begin{pmatrix} \nabla_{\boldsymbol{k}} H_{TSS,00}(\boldsymbol{k}) & \nabla_{\boldsymbol{k}} H_{TSS,01}(\boldsymbol{k}) & 0 \\ \nabla_{\boldsymbol{k}} H_{TSS,10}(\boldsymbol{k}) & \nabla_{\boldsymbol{k}} H_{TSS,11}(\boldsymbol{k}) & -iv_0 \boldsymbol{e}_p \\ 0 & iv_0 \boldsymbol{e}_p & \nabla_{\boldsymbol{k}} \epsilon_R(\boldsymbol{k}) \end{pmatrix}.$$
(16)

Here, $\boldsymbol{e}_p$ denotes the polarization of the pump pulse, and $\boldsymbol{v}_0$ describes the transition dipole strength. The time-dependent Hamiltonian reads

$$H(\boldsymbol{k}, t) = H_{3B}(\boldsymbol{k}) - q\boldsymbol{A}(t) \cdot \boldsymbol{v}_{3B}(\boldsymbol{k}) + \frac{q^2 \boldsymbol{A}(t)^2}{2m}.$$
(17)

Since electron-electron (e-e) scattering processes are expected to occur in the bulk bands (no topological restriction of the scattering phase), we incorporate e-e interactions explicitly for the extended model. In the framework of the nonequilibrium Green's function approach, interactions are incorporated by a self-energy $\Sigma(t, t')$. We construct the simplest possible self-energy approximation that yields generic scattering processes, known as the local second-Born approximation (2Bloc). Within the 2Bloc, the self-energy for each band α is given by

$$\Sigma_\alpha^{\mathrm{e-e}}(t, t') = U_\alpha^2 G_{\alpha\alpha}^{\mathrm{loc}}(t, t') G_{\alpha\alpha}^{\mathrm{loc}}(t, t') G_{\alpha\alpha}^{\mathrm{loc}}(t', t). \tag{18}$$

Here we chose a diagonal self-energy with intra-band interaction $U_\alpha$. The main ingredient for the self-energy (18) is the local Green's function

$$G_{\alpha\alpha}^{\mathrm{loc}}(t, t') = \frac{1}{N_k} \sum_{\boldsymbol{k}} G_{\alpha\alpha}(\boldsymbol{k}, t, t'), \tag{19}$$

where we sum over the corresponding path in momentum space discretized into $N_k$ points. The 2Bloc approximation effectively incorporates the fact that the system is embedded into a higher-dimensional space[57], which is convenient for restricting the dynamics to the 1D cut in the Brillouin zone as done for all other simulations. We chose $U_\alpha = 0$ for the TSSs, ensuring consistency with the calculations performed with the $\boldsymbol{k} \cdot \boldsymbol{p}$ model. For the bulk states we varied $U_\alpha$, but obtained spectra that are in good qualitative agreement with the experiments for a range of values. We chose the realistic value of $U_\alpha = 0.2$ eV for all results presented in the text.

We also include dissipation in all states in the region of the projected bulk band structure. Physically, the main source of this dissipation are propagation effects: electrons excited into bulk bands can freely propagate into the bulk, which diminishes their weight near the surface. Due to the surface sensitivity of ARPES, these electrons cannot be detected, leading to a suppression of the time-resolved ARPES intensity from these bands. This effect can be modelled by including an additional embedding self-energy

$$\Sigma_\alpha^{\mathrm{em}}(t, t') = \gamma_\alpha^2 g_b(t, t'), \tag{20}$$

where $\gamma_\alpha$ describes the coupling strength and where $g_b(t, t')$ represents the noninteracting Green's function with a semi-circular density of states centered at the maximum of the density of states of the projected bulk band structure. We chose $\gamma_\alpha = 10^{-3}$ a.u. for the TSSs, and $\gamma_\alpha = 5 \times 10^4$ a.u. for the bulk, which yields good agreement with the experimental spectra. The embedding self-energy also broadens the bands, thus correctly incorporating a continuum of bulk states into the calculations.

To obtain the time-dependent density matrix, we employ the GKBA equation of motion

$$\frac{d}{dt}\rho(\boldsymbol{k}, t) = -i[H(\boldsymbol{k}, t), \rho(\boldsymbol{k}, t)] + D(\boldsymbol{k}, t), \tag{21}$$

where $D(\boldsymbol{k}, t)$ is the collision integral

$$D(\boldsymbol{k}, t) = -\int_0^t d\bar{t}\left[\Sigma^>(t, \bar{t})G^<(\boldsymbol{k}, \bar{t}, t) - \Sigma^<(t, \bar{t})G^>(\boldsymbol{k}, \bar{t}, t)\right] + h.c. \; . \tag{22}$$

Here the total self-energy $\Sigma(t, t') = \Sigma^{e-e}(t, t') + \Sigma^{em}(t, t')$ enters the collision term. Standard adiabatic switching of the interaction is used to build in the interaction smoothly in real time[58]. After obtaining the time-dependent density matrix, we compute the Green's function (11) in a similar fashion as for the surface-state model and finally obtain the time-resolved ARPES intensity (12).

As shown by our extensive simulations, the TSS dynamics are dominated by the dissipation-free photodressing. In the case of the bulk states, transport from the surface to the bulk region prevents the system from approaching thermal equilibrium on the 100-fs time scale. Experiment and simulation with this basic thermalization model indeed show a good qualitative agreement (Figs. 4e and 4f), supporting the assumption that dynamical surface-bulk coupling dominates relaxation in this system.

**Systematic lightwave ARPES at higher MIR frequencies.** MIR driving fields with higher centre frequencies can induce larger Floquet sideband splitting. This should eventually allow for more efficient coupling between the bulk conduction band and Floquet-Bloch sidebands with smaller Floquet indices. This situation is indeed seen in our systematic lightwave ARPES measurement performed with $\Omega_{MIR}/2\pi = 29$ THz, $31$ THz, and $41$ THz (Extended Data Fig. 10). Whereas the first two cases qualitatively reproduce our observation with the 25 THz driving field, the population of the bulk state after the THz transient becomes slightly stronger for $\Omega_{MIR}/2\pi = 31$ THz despite its lower field strength (0.5 MV cm$^{-1}$). Upon further increase of the driving frequency to 41 THz, the bulk signature is drastically enhanced even though the peak field is further reduced to 0.4 MV cm$^{-1}$. This behaviour can be readily understood by focusing on the overlap between the bulk projection and the second Floquet-Bloch sideband. For photon energies exceeding 0.14 eV, direct optical one-photon transitions are also expected to become possible at the optical band gap located at a finite momentum of ~0.2 Å$^{-1}$ along the $\overline{\Gamma M}$ direction. Yet, for such bulk carriers to contribute to the ARPES signal around the $\overline{\Gamma}$ point they would need to be efficiently accelerated to the centre of the Brillouin zone. For the peak field used here, this process requires a traveling time larger than the oscillation period of the MIR carrier field at $\Omega_{MIR}/2\pi = 41$ THz. Hence, the instantaneous excitation of the bulk populations can be attributed to the more efficient Floquet-bulk coupling at high driving frequencies. We also note that the sideband signature at 41 THz gets weakened already at $t = 10$ fs, demonstrating stronger dissipation from the trivial bulk states and the accelerated decay of Floquet-Bloch states.

Thus, the present lightwave ARPES experiment provides complementary information to HHG spectra recorded with the same frequency range in this system[28]. Whereas HHG detects strongly nonlinear polarization originating from the cusp-like dispersion of the Dirac band, it cannot image the lightwave-driven excursion of carriers in the momentum space and the emergence of Floquet-Bloch replica bands. Conversely, the above observation of bulk populations does not directly mean that the HHG is dominated by the bulk carriers. The electrons scattered into the bulk are actually not as efficient in HHG as the surface Dirac fermions because of their parabolic band structure with an almost harmonic equation of motion. Consequently, HHG at driving frequencies lower than the direct band gap, even with incoming field strengths up to 25 MV cm$^{-1}$, is still dominated by the TSS[28]. In a more general sense, the Floquet-aided interband transition is expected to play a substantial role for HHG in materials with a finite bandgap, where intraband currents are only possible with simultaneous interband excitations.

**Data and code availability** Source data and program codes supporting the findings of this study are available at the Regensburg publication server https://epub.uni-regensburg.de/.

**Method references**

41. Sell, A., Leitenstorfer, A. & Huber, R. Phase-locked generation and field-resolved detection of widely tunable terahertz pulses with amplitudes exceeding 100 MV/cm. *Opt. Lett.* **33**, 2767-2769 (2008).

42. Lu, C.H. *et al.*, Generation of intense supercontinuum in condensed media. *Optica*, **1**, 400-406 (2014).

43. Kokh, K. A. *et al.* Melt growth of bulk $Bi_2Te_3$ crystals with a natural p-n junction. *Cryst. Eng. Comm.* **16**, 581-584 (2014).

44. Giannozzi, P. *et al*. QUANTUM ESPRESSO: A modular and open-source software project for quantum simulations of materials. *J. Phys. Condens. Matter* **21**, 395502 (2009).

45. Nakajima, S. The crystal structure of $Bi_2Te_{3-x}Se_x$. *J. Phys. Chem. Solids* **24**, 479-485 (1963).

46. Zhang, P. *et al.* A precise method for visualizing dispersive features in image plots. Rev. Sci. Instrum. **82**, 043712 (2011).

47. Keunecke, M. *et al.* Electromagnetic dressing of the electron energy spectrum of Au(111) at high momenta. *Phys. Rev. B* **102**, 161403(R) (2020).

48. Farrell, A., Arsenault, A., & Pereg-Barnea, T., Dirac cones, Floquet side bands, and theory of time-resolved angle-resolved photoemission. *Phys. Rev. B* **94**, 155304 (2016).

49. Liu, C.X. *et al*. Model Hamiltonian for topological insulators. *Phys. Rev. B* **82**, 045122 (2010).

50. Schüler, M., Marks, J.A., Murakami, Y., Jia, C. & Devereaux, T.P. Gauge invariance of light-matter interactions in first-principle tight-binding models. *Phys. Rev. B* **103**, 155409 (2021).

51. Sato, S.A. *et al*. Microscopic theory for the light-induced anomalous Hall effect in graphene. *Phys. Rev. B* **99**, 214302 (2019).

52. Floss, I. *et al*. *Ab initio* multiscale simulation of high-order harmonic generation in solids, *Phys. Rev. A* **97**, 011401(R) (2018).

53. Freericks, J.K., Krishnamurthy, H.R. & Pruschke, Th. Theoretical Description of Time-Resolved Photoemission Spectroscopy: Application to Pump-Probe Experiments. *Phys. Rev. Lett.* **102**, 136401 (2009).

54. Lipavský, P., Špička, V. & Velický, B. Generalized Kadanoff-Baym ansatz for deriving quantum transport equations, *Phys. Rev. B* **34**, 6933 (1986).

55. Schüler, M. & Sentef, M.A. Theory of subcycle time-resolved photoemission: Application to terahertz photodressing in graphene. *J. Electron Spectrosc. Relat. Phenom.* **253**, 147121 (2021).

56. Yazyev, O.V., Morre, J.E. & Louie, S.G. Spin Polarization and Transport of Surface States in the Topological Insulators $Bi_2Se_3$ and $Bi_2Te_3$ from First Principles. *Phys. Rev. Lett.* **105**, 266806 (2010).

57. Schüler, M., Murakami, Y., & Werner, P. Nonthermal switching of charge order: Dynamical slowing down and optimal control. *Phys. Rev. B* **97**, 155136 (2018).

58. Balzer, K., & Bonitz, M. *Nonequilibrium Green's Functions Approach to Inhomogeneous Systems*, Ch. 2 (Springer, 2013).

**Acknowledgements** We thank Klaus Richter and Zhensheng Tao for helpful discussions. The work in Marburg has been supported by the Deutsche Forschungsgemeinschaft (DFG, German Research Foundation) through Project ID 223848855-SFB 1083 and Research Grant GU 495/2-2 and HO 2295/9. M.S. thanks the Swiss National Science Foundation SNF for its support with an Ambizione grant (project No. 193527). The work in Regensburg has been supported by the Deutsche Forschungsgemeinschaft (DFG) through Project ID 422 314695032-SFB 1277 (Subproject A05) as well as Research Grant HU1598/8. O.E.T. and K.A.K. have been supported by the RFBR and DFG (project No.21-52-12024) in the part of crystal growth and Russian Science Foundation (project No. 22-12-20024) in the part of crystal characterization and state contract of IGM SB RAS and ISP SB RAS. S.I. acknowledges support from JSPS Postdoctoral Fellowship for Research Abroad. M.A.S. acknowledges support by the Deutsche Forschungsgemeinschaft (DFG) through the Emmy Noether Programme (SE 2558/2).

**Author Contributions** S.I., J.G., U.H., and R.H. conceived the experiment. S.I., M.M., S.S., J.F., J.R., D.A., and J.G. carried out the experiment. K.A.K. and O.E.T. grew the crystals and characterized their properties. M.S. and M.A.S. developed and carried out the quantum nonequilibrium calculations. S.I. carried out the DFT calculations and developed the minimal quantum model. All authors analysed the data and discussed the results. S.I., U.H., and R.H. wrote the manuscript with contributions from all authors.

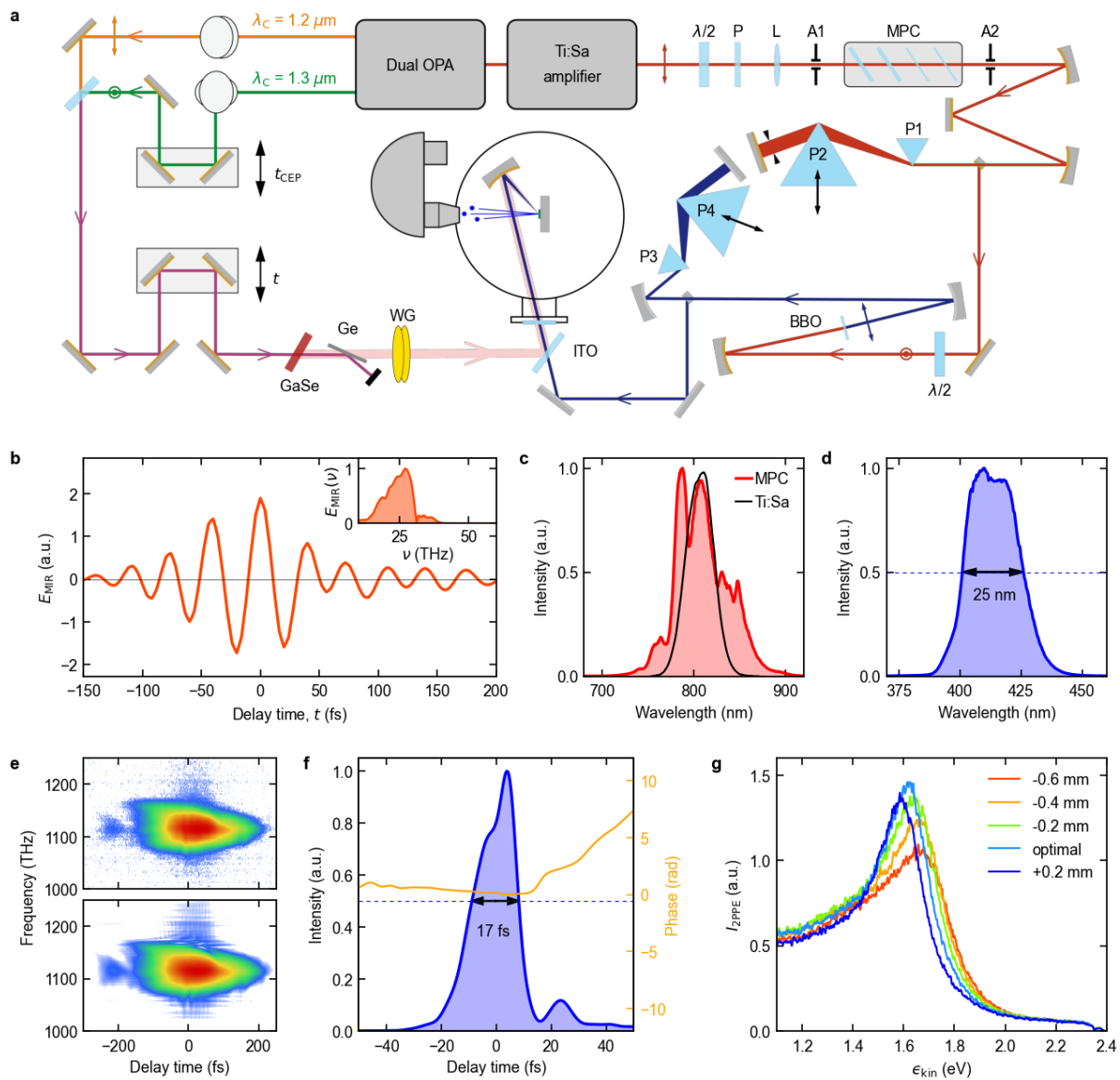

**Extended Data Fig. 1 | Lightwave ARPES setup. a**, The output of a titanium-sapphire (Ti:Sa) amplifier system is used to generate intense MIR pulses with stable CEP by DFG of the near-infrared pulse trains from a dual-branch OPA (left half). In parallel, broadband UV probe pulses are generated (right half), which are spatially overlapped with the MIR field transients and guided into the UHV chamber for ARPES. Arrows indicate the polarization state of each beam. **b**, MIR waveform measured by electro-optic sampling. Inset: corresponding amplitude spectrum. **c**, Spectral intensity of the laser fundamental (black) and the MPC (red). **d**, Broadband UV spectrum (FWHM, 25 nm). **e**, Measured (top) and reconstructed (bottom) XFROG traces of the UV probe pulse. **f**, Intensity and phase profiles of the reconstructed UV pulse providing a duration of 17 fs (FWHM). **g**, Dependence of the 2PPE

spectra on the position of prism P4. The maximum 2PPE intensity is reached when the UV pulses in the UHV chamber are optimally compressed.

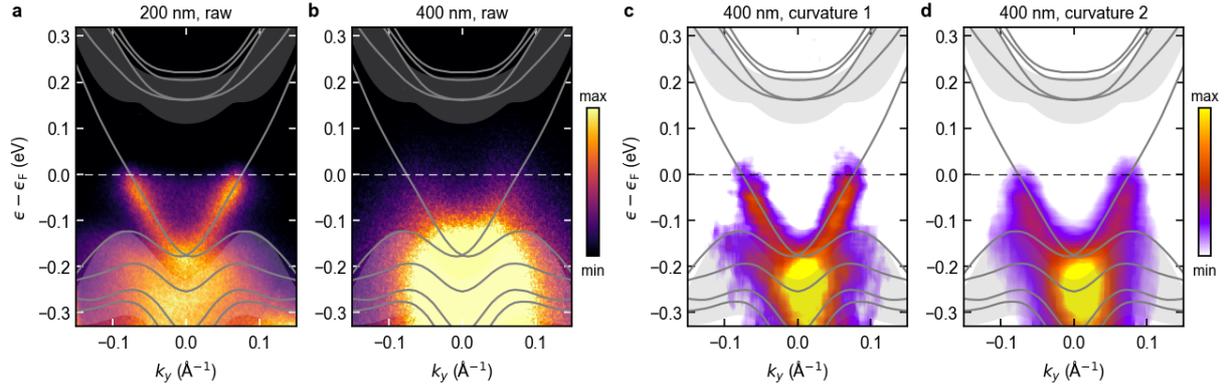

**Extended Data Fig. 2 | Validation of broadband 2PPE probing and curvature image processing.**
**a**, One-photon ARPES map of Bi$_2$Te$_3$ measured with 200-nm probe light with a bandwidth of 2 nm.
Grey curves and shaded areas depict the DFT band structure of Bi$_2$Te$_3$. **b**, 2PPE-ARPES map of Bi$_2$Te$_3$
measured with the broadband 400-nm probe. **c**, **d**, Curvature image filtering of the 2PPE spectrum of **b**
precisely retrieves the band structure obtained with one-photon ARPES in **a**. The band visibility is
controlled by the factor $C_0$ in Eq. (1) and the level of the preceding smoothing process. Panels **c** and **d**
compare 2D curvature filtered results obtained with a preliminary smoothing filter size of 0.035
Å⁻¹×0.06 eV and 0.05 Å⁻¹×0.09 eV, respectively.

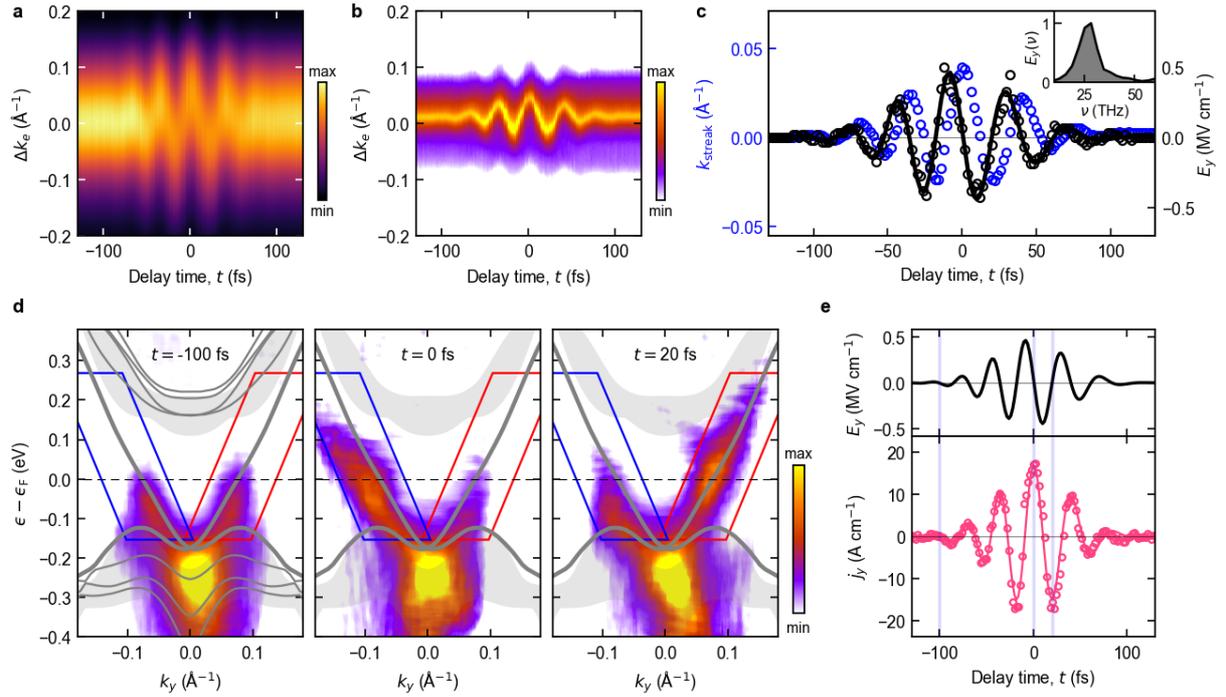

**Extended Data Fig. 3 | Reconstruction of electric-field and current waveforms. a**, Momentum-streaking trace obtained by integrating raw ARPES spectra in the region of the bulk valence band. **b**, Curvature-filtered momentum-streaking trace of the same data. **c**, Momentum-streaking waveform extracted from **b** (blue) and resulting s-polarized electric-field waveform (black). Open circles indicate data points, the solid line is a fit to the analytical waveform, see Eq. (2). Inset: amplitude spectra of the Fourier-transformed field waveform. **d**, Curvature-filtered lightwave ARPES maps measured at $t = -100$ fs, 0 fs, and 20 fs (vertical lines in **e**). Blue and red boxes indicate areas where the electron occupation is integrated for the determination of surface currents. **e**, Upper panel: electric field waveform from (**c**). Lower panel: current-density waveform extracted from ARPES data (red circles) and calculated from the measured electric field by means of the semiclassical Boltzmann equation (red line).

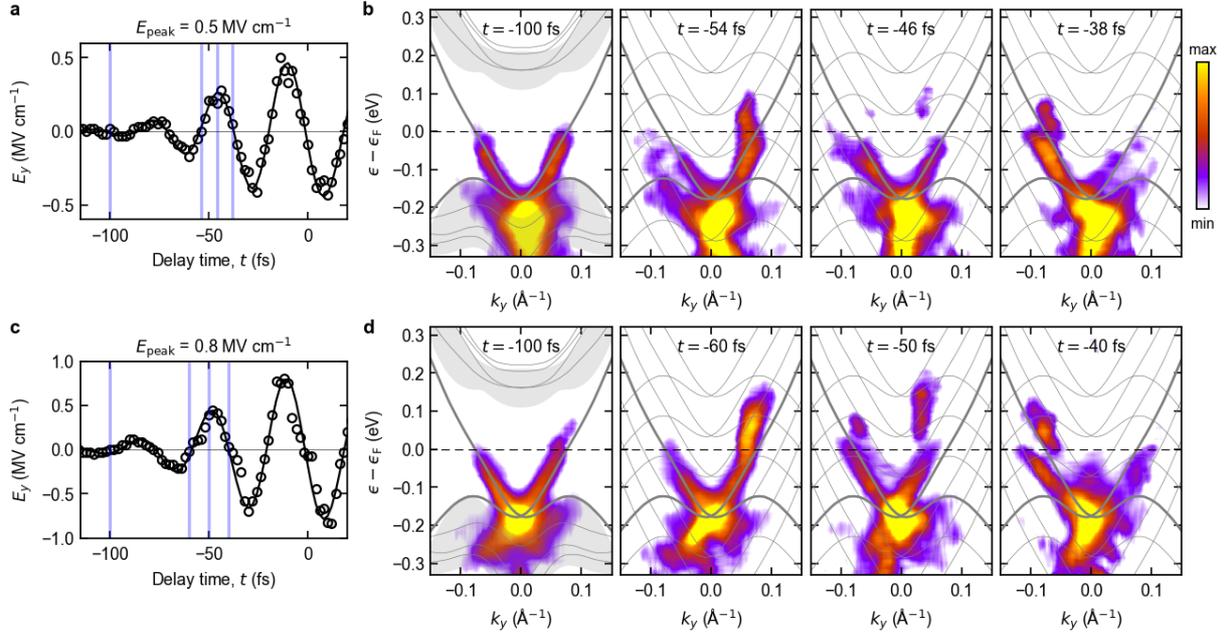

**Extended Data Fig. 4 | THz field-strength dependence of Floquet sidebands. a**, Measured MIR driving field with a centre frequency of 25 THz and a peak field strength of 0.5 MV cm⁻¹. Circles, experimental data from streaking reconstruction, solid line: numerical fit of Eq. (2). **b**, Curvature-filtered lightwave ARPES maps at $t$ = -100 fs, -54 fs, -46 fs, and -38 fs. These temporal positions are indicated in **a**. **c**, **d**, Comparable data set for a peak field strength of 0.8 MV cm⁻¹ (data set shown in Fig. 2).

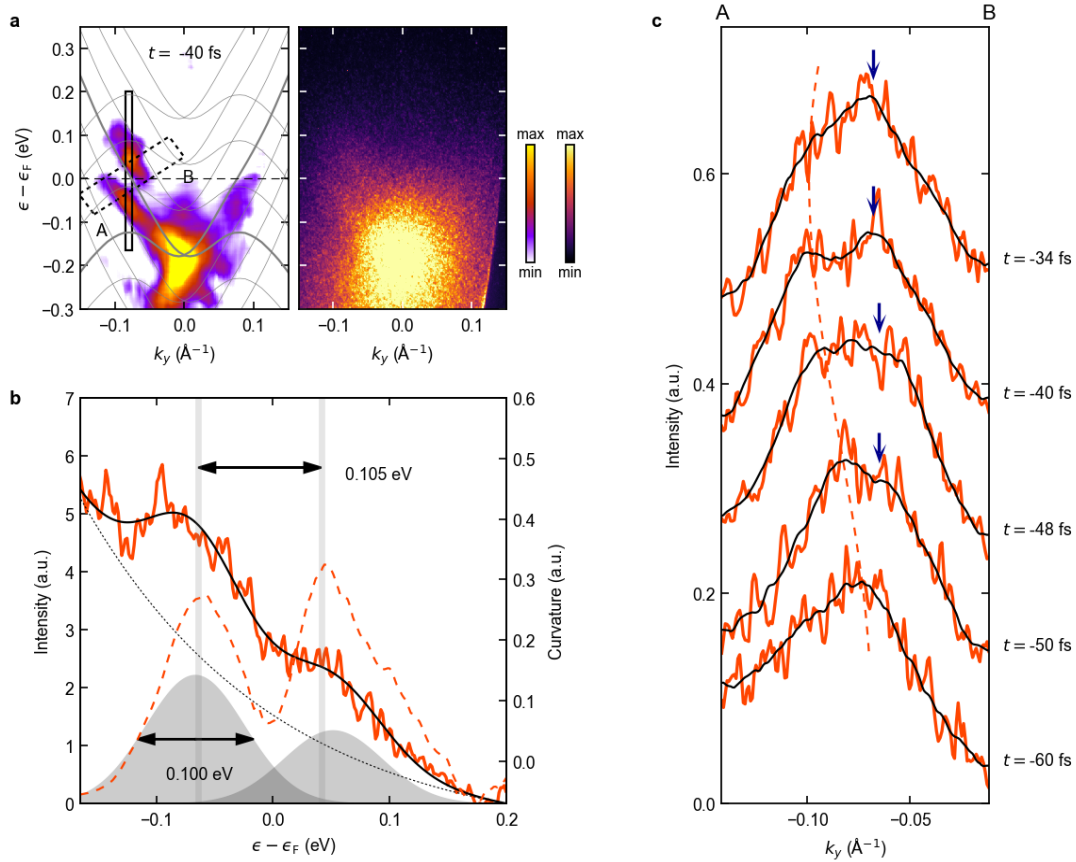

**Extended Data Fig. 5 | Sideband signatures in raw ARPES maps. a**, Left, the curvature-filtered lightwave ARPES map at $t$ = -40 fs adapted from Fig. 2. Right, the corresponding raw ARPES image. The solid and dashed boxes depict the regions considered to extract the energy distribution curves in **b** and the one-dimensional (1D) intensity distribution curves in **c**, respectively. **b**, An energy distribution curve obtained from the raw ARPES map (orange solid line) and its fit (black solid line) by using two Gaussian peaks (grey shaded areas) and an exponential background (black dotted line). The orange dashed line shows an energy distribution curve extracted from the curvature image, and the vertical grey lines highlight energy splitting, which is quantitatively consistent with the driving frequency of 25 THz. **c**, 1D intensity distribution curves in the direction perpendicular to the left Dirac branch (indicated by AB in **a**) from raw ARPES data measured at $t$ = -60 fs, -50 fs, -48 fs, -40 fs, and -34 fs. The curves are normalized and vertically offset for clarity. The orange lines show raw intensity curves averaged inside the extraction region, and the black lines are smoothed by a 1D filter. Blue arrows highlight positions of emergent peaks in the raw data, and the dashed orange line is a guide tracking the evolution of the ground-state peak.

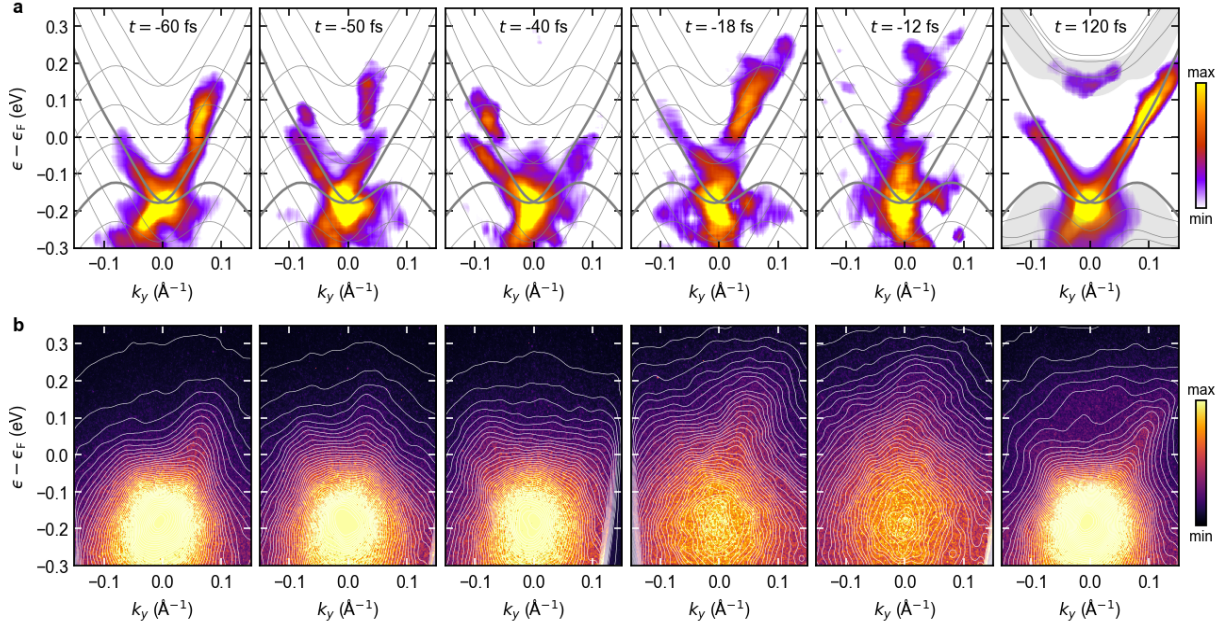

**Extended Data Fig. 6 | Systematic comparison of curvature-filtered and raw ARPES spectra. a,** The curvature-filtered lightwave ARPES maps measured at $t$ = -60, -50, -40, -18, -12, and 120 fs adapted from Fig. 2 and Fig. 4. **b,** The corresponding raw ARPES images. Overlaid white lines depict 2D contours of the ARPES intensities. Formation of additional peaks in regions where there is no equilibrium band dispersion is clearly visible in the raw spectra. In addition, strong suppression of the bulk intensity, which spreads into Floquet-Bloch bands, is directly observed in the intensity distributions especially at $t$ = -18 fs and $t$ = -12 fs. After the driving field leaves the surface at $t$ = 120 fs, the system basically recovers the original spectral feature, except for the persistent intensities in the bulk conduction band and the enhanced occupation in the surface states due to finite heating.

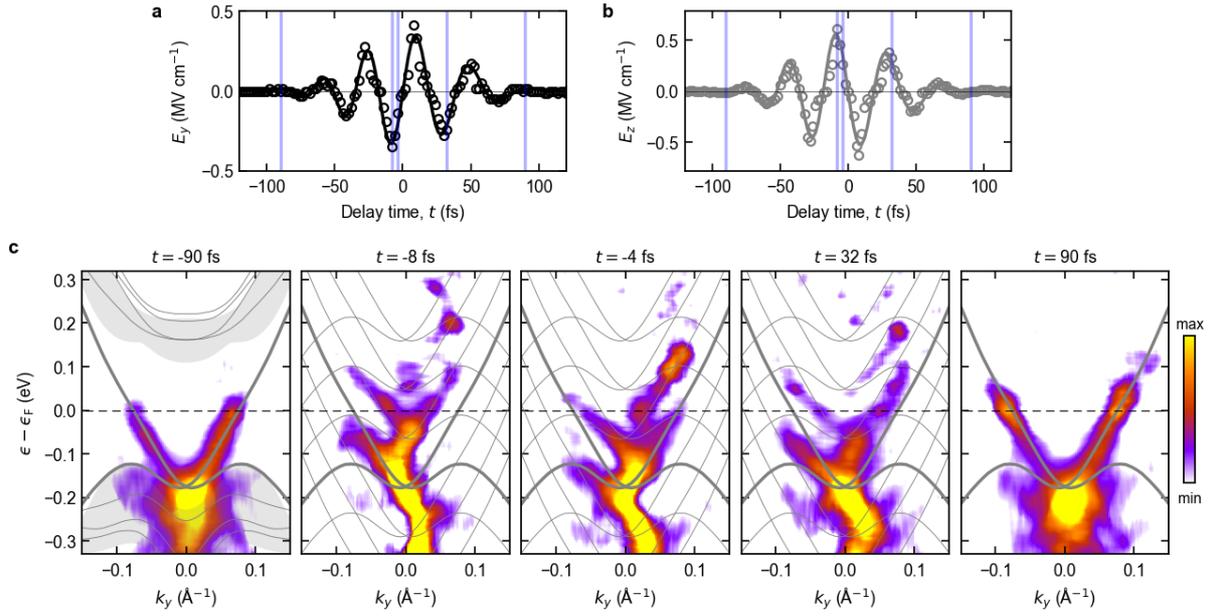

**Extended Data Fig. 7 | Observation of Floquet-Volkov sidebands with mixed s- and p-polarized electric fields. a**, s-polarized and **b**, p-polarized MIR electric field waveform (centre frequency, 25 THz) reconstructed from momentum and energy streaking of a single lightwave ARPES measurement, respectively. **c**, Streaking-compensated, curvature-filtered lightwave ARPES maps measured at $t$ = -90 fs, -8 fs, -4 fs, 32 fs, and 90 fs (vertical lines in **a** and **b**) with the driving field characterized in **a** and **b**, featuring both s- and p-polarization components.

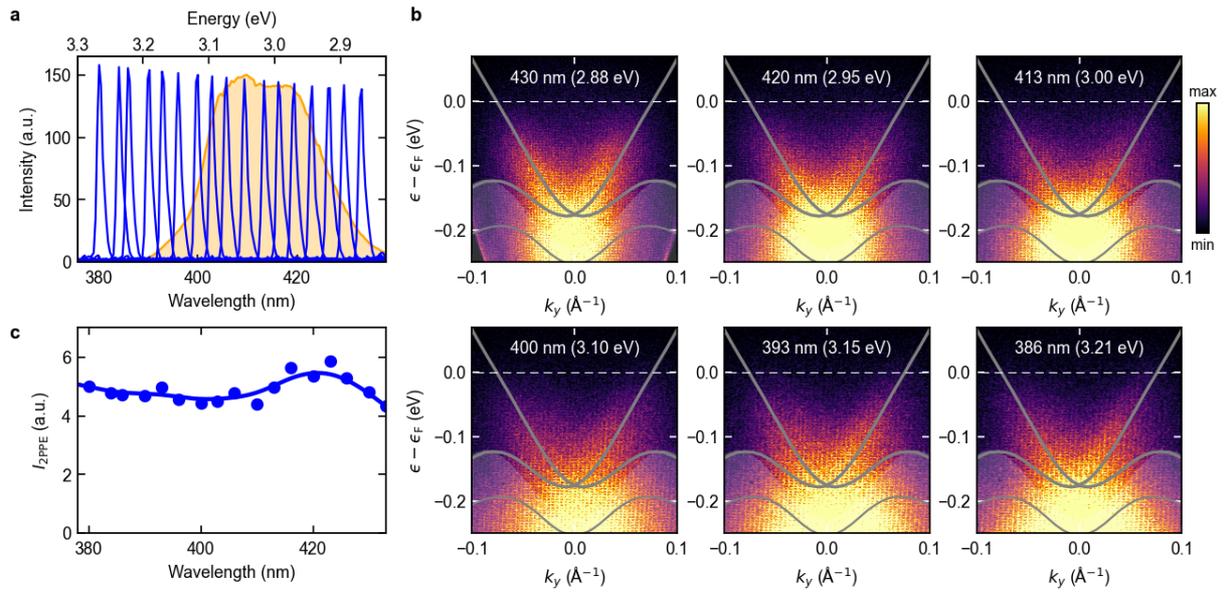

**Extended Data Fig. 8 | Absence of resonance effects in 2PPE with broadband 400-nm probe pulses**
**a**, Spectra of a picosecond, wavelength-tunable light source (blue curves) and the broadband UV probe used for lightwave ARPES (orange). The intensity of the blue spectra was adjusted for constant photon flux. **b**, 2PPE-ARPES maps measured with the tunable probe for different wavelengths (photon energies) as indicated in the respective panels. **c**, Integrated 2PPE intensity, $I_{2PPE}$, as a function of the probe centre wavelength. The solid curve is a guide to the eye.

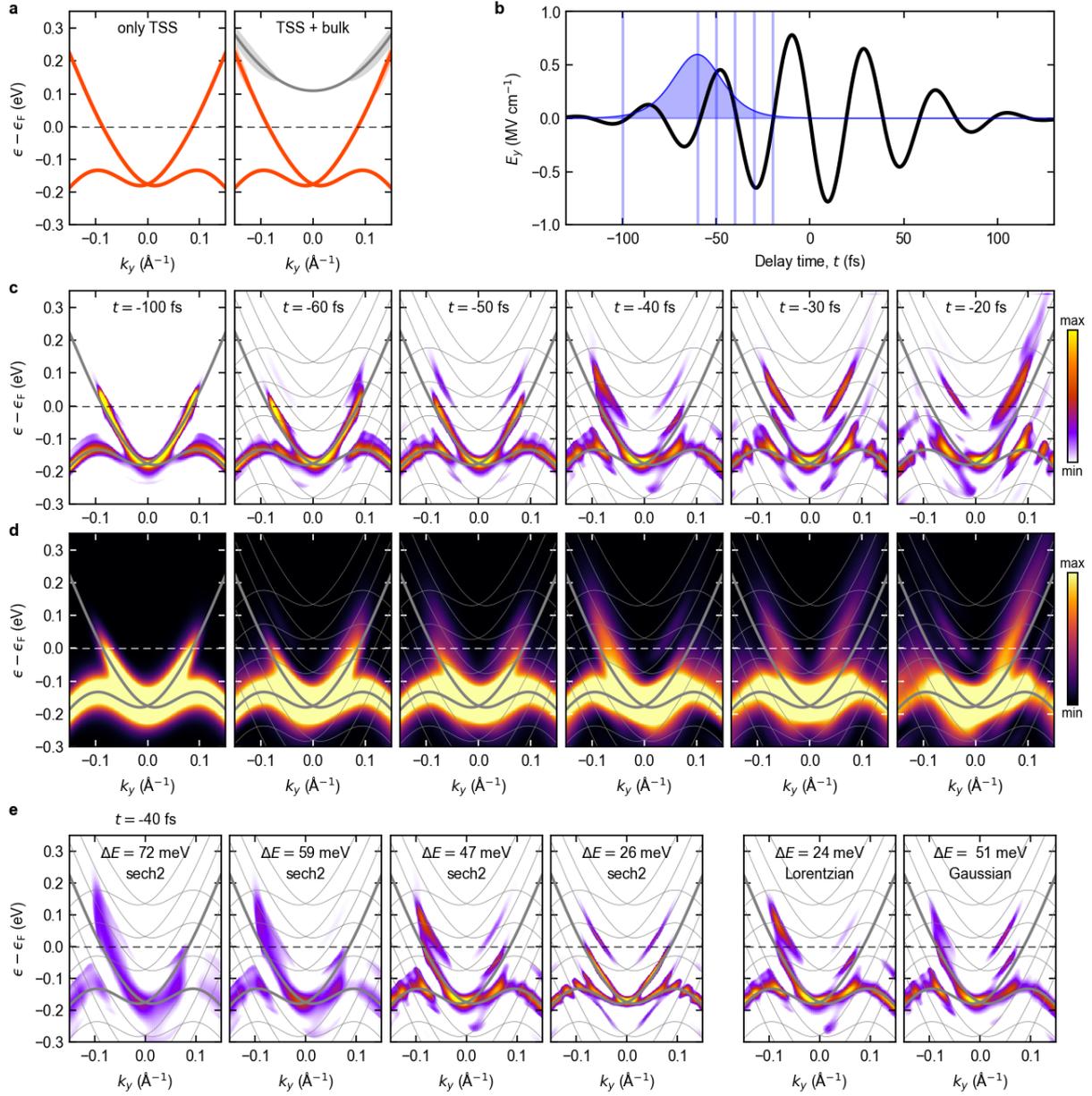

**Extended Data Fig. 9 | Model band structure of Bi₂Te₃ and time-resolved ARPES simulations. a**, Band structure used in our lightwave ARPES calculations. Left, the pure TSS band structure given by the $\boldsymbol{k} \cdot \boldsymbol{p}$ Hamiltonian (Eq. 7). Right, the 3B model used to describe dynamical Floquet-bulk coupling. The grey shaded area illustrates delocalization of the bulk states to account for self-energy effects in our model (see Methods). **b**, Input driving waveform which well reproduces the experimental result in Fig. 2b (black curve) and sech-shaped probe pulses with a bandwidth of 47 meV (intensity FWHM) (filled blue curve). Thin vertical lines indicate the temporal positions shown in **c**. **c**, Curvature-filtered ARPES images for the 47 meV probe calculated at different time steps with the pure TSS band structure (red curve in **a**, see also Figs. 2 and 4 for corresponding experimental data). **d**, Corresponding raw

ARPES spectra. **e**, ARPES spectra for different bandwidths and shapes of the probe pulse at a fixed delay time $t$ = -40 fs.

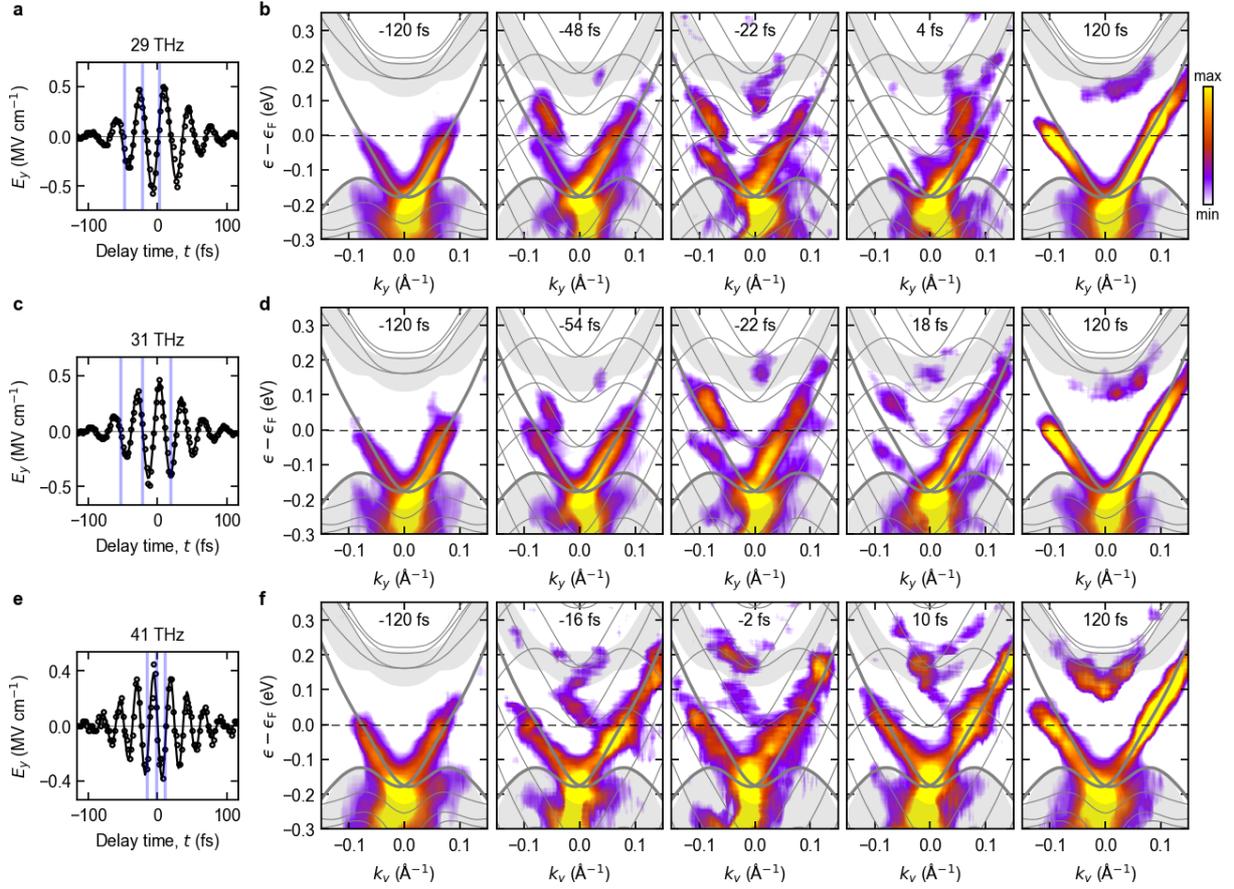

**Extended Data Fig. 10 | Systematic lightwave ARPES at MIR frequencies. a**, MIR waveform centred at 29 THz. Circles represent the reconstruction from momentum streaking, the solid curve is a numerical fit of Eq. (2). **b**, Curvature-filtered lightwave ARPES maps measured before the arrival of the THz field ($t$ = -120 fs), at three temporal positions during the field transient (vertical lines in **a**), and after the THz field has left the surface ($t$ = 120 fs). **c**, **d**, Corresponding data set obtained for a centre frequency of 31 THz. **e**, **f**, Corresponding data set for a centre frequency of 41 THz. The larger splitting at 41 THz (photon energy, 0.16 eV) facilitates the coupling between Floquet sidebands and the bulk conduction band already at lower field strengths than in the case of excitation at 25 THz (Fig. 4).